\documentclass[twocolumn,astrosymb]{aastex631}

\newcommand{\siii}{Si\,{\sc ii}\,}

\newcommand{\civ}{C\,{\sc iv}\,}

\newcommand{\oiii}{[O\,{\sc iii]}\,}
\newcommand{\oii}{[O\,{\sc ii]}\,}

\newcommand{\mgii}{Mg\,{\sc ii}\,}

\newcommand{\ha}{H$\alpha$\,}

\newcommand{\hdelta}{H$\delta$\,}

\newcommand{\caii}{Ca\,{\sc ii}\,}
\newcommand{\mgi}{Mg\,{\sc i}\,}

\def\sersic{S\'{e}rsic\,}

\usepackage{physunits}
\usepackage{subfigure}
\usepackage{amsmath}
\usepackage{xcolor}
\newcommand{\myrev}[1]{#1}

\begin{document}

\title{Quiescent Host Galaxies of Extended Quasars Revealed by Spectrophotometric Decomposition}

\author[0000-0002-1234-552X]{Shengxiu Sun}
\author[0000-0003-4176-6486]{Linhua Jiang}
\author[0000-0003-0230-6436]{Zhiwei Pan}
\affiliation{Department of Astronomy, School of Physics, Peking University, Beijing 100871, China}
\affiliation{Kavli Institute for Astronomy and Astrophysics, Peking University, Beijing 100871, China}

\author[0000-0002-2949-2155]{Ma\l gorzata Siudek}
\affiliation{Instituto de Astrof\'{\i}sica de Canarias, V\'{\i}a L\'actea, 38205 La Laguna, Tenerife, Spain}
\affiliation{Instituto de Astrof\'isica de Canarias (IAC); Departamento de Astrof\'isica, Universidad de La Laguna (ULL), 38200, La Laguna, Tenerife, Spain}

\author[0000-0003-4440-259X]{Mar Mezcua}
\affiliation{Institute of Space Sciences, ICE-CSIC, Campus UAB, Carrer de Can Magrans s/n, 08913 Bellaterra, Barcelona, Spain}
\affiliation{Institut d'Estudis Espacials de Catalunya (IEEC), c/ Esteve Terradas 1, Edifici RDIT, Campus PMT-UPC, 08860 Castelldefels, Spain}
\author[0000-0003-3292-8631]{Gaocheng Yin}
\affiliation{Department of Astronomy, School of Physics, Peking University, Beijing 100871, China}
\affiliation{Kavli Institute for Astronomy and Astrophysics, Peking University, Beijing 100871, China}

\author[0000-0002-5854-7426]{Swayamtrupta Panda}
\thanks{Gemini Science Fellow}
\affiliation{International Gemini Observatory/NSF NOIRLab, Casilla 603, La Serena, Chile}

\author[0000-0001-9457-0589]{Wei-Jian Guo}
\affiliation{Key Laboratory of Optical Astronomy, National Astronomical Observatories, Chinese Academy of Sciences, Beijing 100012, China}

\author[0000-0001-6098-7247]{Steven Ahlen}
\affiliation{Department of Physics, Boston University, 590 Commonwealth Avenue, Boston, MA 02215, USA}
\author{David Brooks}
\affiliation{Department of Physics \& Astronomy, University College London, Gower Street, London, WC1E 6BT, UK}
\author{Todd Claybaugh}
\affiliation{Lawrence Berkeley National Laboratory, 1 Cyclotron Road, Berkeley, CA 94720, USA}
\author[0000-0002-1769-1640]{Axel de la Macorra}
\affiliation{Instituto de F\'{\i}sica, Universidad Nacional Aut\'{o}noma de M\'{e}xico,  Circuito de la Investigaci\'{o}n Cient\'{\i}fica, Ciudad Universitaria, Cd. de M\'{e}xico  C.~P.~04510,  M\'{e}xico}
\author{Peter Doel}
\affiliation{Department of Physics \& Astronomy, University College London, Gower Street, London, WC1E 6BT, UK}
\author[0000-0001-9632-0815]{Enrique Gaztañaga}
\affiliation{Institut d'Estudis Espacials de Catalunya (IEEC), c/ Esteve Terradas 1, Edifici RDIT, Campus PMT-UPC, 08860 Castelldefels, Spain}
\affiliation{Institute of Cosmology and Gravitation, University of Portsmouth, Dennis Sciama Building, Portsmouth, PO1 3FX, UK}
\affiliation{Institute of Space Sciences, ICE-CSIC, Campus UAB, Carrer de Can Magrans s/n, 08913 Bellaterra, Barcelona, Spain}
\author{Gaston Gutierrez}
\affiliation{Fermi National Accelerator Laboratory, PO Box 500, Batavia, IL 60510, USA}
\author[0000-0003-3510-7134]{Theodore Kisner}
\affiliation{Lawrence Berkeley National Laboratory, 1 Cyclotron Road, Berkeley, CA 94720, USA}
\author{Andrew Lambert}
\affiliation{Lawrence Berkeley National Laboratory, 1 Cyclotron Road, Berkeley, CA 94720, USA}
\author[0000-0003-1838-8528]{Martin Landriau}
\affiliation{Lawrence Berkeley National Laboratory, 1 Cyclotron Road, Berkeley, CA 94720, USA}
\author[0000-0002-1125-7384]{Aaron Meisner}
\affiliation{NSF NOIRLab, 950 N. Cherry Ave., Tucson, AZ 85719, USA}
\author{Ramon Miquel}
\affiliation{Instituci\'{o} Catalana de Recerca i Estudis Avan\c{c}ats, Passeig de Llu\'{\i}s Companys, 23, 08010 Barcelona, Spain}
\affiliation{Institut de F\'{i}sica d’Altes Energies (IFAE), The Barcelona Institute of Science and Technology, Edifici Cn, Campus UAB, 08193, Bellaterra (Barcelona), Spain}
\author[0000-0002-2733-4559]{John Moustakas}
\affiliation{Department of Physics and Astronomy, Siena College, 515 Loudon Road, Loudonville, NY 12211, USA}
\author[0000-0001-6979-0125]{Ignasi P\'erez-R\`afols}
\affiliation{Departament de F\'isica, EEBE, Universitat Polit\`ecnica de Catalunya, c/Eduard Maristany 10, 08930 Barcelona, Spain}
\author[0000-0002-9646-8198]{Eusebio Sanchez}
\affiliation{CIEMAT, Avenida Complutense 40, E-28040 Madrid, Spain}
\author{David Schlegel}
\affiliation{Lawrence Berkeley National Laboratory, 1 Cyclotron Road, Berkeley, CA 94720, USA}
\author{Michael Schubnell}
\affiliation{Department of Physics, University of Michigan, 450 Church Street, Ann Arbor, MI 48109, USA}
\affiliation{University of Michigan, 500 S. State Street, Ann Arbor, MI 48109, USA}
\author[0000-0002-6588-3508]{Hee-Jong Seo}
\affiliation{Department of Physics \& Astronomy, Ohio University, 139 University Terrace, Athens, OH 45701, USA}
\author{David Sprayberry}
\affiliation{NSF NOIRLab, 950 N. Cherry Ave., Tucson, AZ 85719, USA}
\author[0000-0003-1704-0781]{Gregory Tarl\'{e}}
\affiliation{University of Michigan, 500 S. State Street, Ann Arbor, MI 48109, USA}
\author{Benjamin Alan Weaver}
\affiliation{NSF NOIRLab, 950 N. Cherry Ave., Tucson, AZ 85719, USA}

\author[0000-0001-5381-4372]{Rongpu Zhou}
\affiliation{Lawrence Berkeley National Laboratory, 1 Cyclotron Road, Berkeley, CA 94720, USA}

\begin{abstract}

Previous works of low-redshift quasar host galaxies have focused on compact quasars and found that their host galaxies are mainly star-forming galaxies. Here we present a study of host galaxies for quasars with extended morphologies in ground-based optical images. We select a sample of more than 1000 type 1 quasars at redshift $0.1<z<1$ that are classified as extended objects by DESI. Combining high-resolution spectra from DESI and high-quality images from Subaru HSC, we develop a spectrophotometric decomposition technique to iteratively decompose each quasar into an AGN component and its host galaxy. The technique can effectively break the degeneracy between the AGN and host components and capture the host spectral features. Our results show that the host galaxies of most quasars have low star-formation rates (SFRs) and low specific SFRs, indicating that they are quiescent galaxies. Many of them exhibit prominent post-starburst features with the existence of significant old stellar populations. These properties are quite different from the nature of compact quasars with star-forming host galaxies. 
In addition, the relation between the black hole mass and stellar mass for our sample is broadly consistent with the canonical local relations.
This work is complementary to the previous studies and suggests that the host galaxies of low-redshift quasars are more diverse than what was thought.

\end{abstract}


\section{Introduction} \label{sec:intro}

The co-evolution of supermassive black holes (SMBHs) and their host galaxies is one of the main topics in the field of galaxy evolution. Most massive galaxies with stellar bulges at low redshift are thought to harbor SMBHs, and the properties of the SMBHs show tight correlations with those of their host galaxies, indicating a strong connection between the nuclear activity and galaxy evolution \citep[e.g.,][]{2011MNRAS.412.2211G_m-sigma, 2013ARA&A..51..511K}. Not only does the scaling relation exist in inactive galaxies, but also in active galaxies \citep[e.g.,][]{2023NatAs...7.1376Z_evolutionary_paths}.

The physical mechanism that produces such tight relationships is elusive due to the substantial mismatch between the dynamical radius of central SMBHs ($\sim \mathrm{1 \, pc}$) and the size of host galaxies ($\sim \mathrm{1 \, kpc}$). The energetic outflows from active galactic nuclei (AGNs) associated with their activities are thought to have a significant correlation with physical processes in the host galaxy, and many observational phenomena can be explained by the mechanism of the AGN feedback \citep[e.g.,][]{2012ARA&A..50..455F_Fabian}. Cosmological simulations can reproduce the local correlations by invoking the AGN feedback as the physical mechanism for such a strong connection \citep[e.g.,][]{2008ApJS..175..356Hopkins}, or by assuming that the galaxy evolution and SMBH growth are closely linked as they share the same gas reservoir \citep[e.g.,][]{2015ApJ...805L...9Cen}. Others think that such a physical connection may not be necessary, and the statistical convergence from galaxy assembly alone (mergers) may be able to reproduce the observed correlations \citep[e.g.,][]{2007ApJ...671.1098Peng}. Therefore, the intrinsic origin of this co-evolution and the prevalence of the AGN feedback scenario need further explorations.

The key is to acquire better understanding of the relations between the black hole masses and the properties of their hosts. It is also of vital importance to study these relations as a function of redshift and other physical parameters such as the AGN environment, and determine how, when, and why the correlations emerge and evolve over cosmic time. Achieving robust measurements of these properties is challenging. On one hand, for quasars whose AGNs are in the most active phase, their brightness often prevents us from directly probing their host galaxies in the optical wavelength ranges \citep[e.g.,][]{Mechtley_2012}. On the other hand, the emission originating from AGNs and their host galaxy components are tightly coupled. Previous studies usually rely on two independent methods to estimate the host galaxy components, including imaging decomposition and spectral decomposition. Results from two different methods were often inconsistent, and information drawn from individual methods was limited. Even in the era of the James Webb Space Telescope JWST \citep[JWST;][]{Gardner_2023}, we still cannot make robust detections of  host galaxies for many quasars \citep[e.g.,][]{Stone_2024}. It has been found that high-$z$ quasars exhibit large deviations from the local $M_{\mathrm{BH}} - \sigma_{\star}$ relation, and their $M_{\mathrm{BH}}$ can exceed the traditional relation by more than 2 dex \citep[e.g.,][]{maiolino_2024_overmassive_BHs}. These overmassive SMBHs at high redshift ($z>4$) hint a possible redshift evolution of this scaling relation and heavy black hole seeds \citep[e.g.,][]{2023ApJ...957L...3P-overmassive_bhs}. 

The imaging decomposition of AGN host galaxies is complex, as the point spread function (PSF) must be taken into consideration. Previous works based on Hubble Space Telescope (HST) images in deep survey fields such as the Cosmic Evolution Survey \citep[COSMOS;][]{2007ApJS..172....1S, 2022ApJS..258...11W} and Chandra Deep Field-South \citep[CDFS;][]{2002ApJS..139..369G, 2017ApJS..228....2L} reported no redshift evolution in the $M_{\rm{BH}}-M_{\star}$ relation \citep{2013ApJ...767...13S_ss, 2016ApJ...830..156_Mechtley}.
Many tools have been developed to perform the imaging decomposition, including \texttt{GALFIT} \citep{2002AJ....124..266P_galfit_orgpaper}, \texttt{lenstronomy}  \citep{2018PDU....22..189B_lenstronomy}, and \texttt{galight} \citep{2020ApJ...888...37D_Mbh_Mstar_Ding2020}, a more convenient version of \texttt{lenstronomy}.
\citet{2021ApJ...918...22_Lijunyao} used \texttt{lenstronomy} for the imaging decomposition of Hyper Suprime-Cam Subaru Strategic Program (HSC-SSP) data and measured the structural and photometric properties of the Sloan Digital Sky Survey \citep[SDSS;][]{2000AJ....120.1579Y} quasar host galaxies. They used a PSF and \sersic \,profile to characterize the quasar and host components, respectively. Their found that quasars are preferentially hosted by massive star-forming galaxies with disk-like light profiles, suggesting that galaxies are concurrently fueling their SMBHs and building their stellar bulge from a centrally concentrated gas reservoir. \citet{2023Natur.621...51Ding_host-stellar-light_z6} used \texttt{galight} to model quasar images in a similar manner and to detect host stellar light of quasars during the reionization epoch ($z > 6$). The NIRSpec rest-frame optical spectrum for one of their quasars exhibits prominent stellar absorption features, providing solid evidence for its host galaxy emission.

After the photometric decomposition, broadband spectral energy distribution (SED) modeling can be performed for the decomposed AGN and host galaxy components separately \citep[e.g.,][]{2013ARA&A..51..393C, BC03, Salim_2007, 10.1093/mnras/sty2169, Leja_2019,  2016ApJ...833...98C}. The SED fitting for galaxies estimates their physical parameters, including redshift $z$, star formation rate (SFR), velocity dispersion $\sigma_{\star}$, stellar mass $M_{\star}$, metallicity $Z$, and dust attenuation of the galaxy. These measurements derived purely from the broadband SED fitting are less reliable than those from the spectra with same wavelength coverage. The SED fitting results are largely dependent on the chosen templates and the photometric bands available, and thus often have large uncertainties \citep{Bingjie_RUBIES_SFH_Mstar, 2024A&A...691A.308S}. Nevertheless, the broadband SED modeling is an efficient tool that is easy to accomplish.

\myrev{The spectral decomposition of quasars requires a comprehensive model construction of both AGN and host stellar components. Various efforts have been dedicated to this task. Details can be found in \citet{Ren_2024}. Traditional approaches typically match observed spectra with stellar population synthesis models \citep[e.g.,][]{BC03, 2006MNRAS.371..703S_MILES} using codes like \texttt{STARLIGHT} \citep{2005MNRAS.358..363C, Junjie_CLAGN_host}. While some earlier studies explored data-driven techniques such as principal component analysis \citep[PCA; e.g.,][]{2004AJ....128..585_Yip_PCA}, a long-standing challenge in these 1D spectral methods is the severe mathematical degeneracy between the featureless AGN power-law continuum and the underlying host stellar continuum.}

\myrev{To break this degeneracy, several studies have leveraged high-resolution images as a spatial prior for the spectral decomposition \citep[e.g.,][]{Du_2014}. By performing a 2D morphological decomposition on images to separate the unresolved AGN point source from the extended host galaxy, researchers can quantify the host galaxy starlight. This photometric flux acts as a robust prior constraint, anchoring the amplitude of the stellar templates during the 1D spectral fitting. Advancing this cross-instrument philosophy, tools such as \texttt{GalfitS} \citep[R. Li \& L. C. Ho, in preparation;][]{2025ApJ...983...60C} have recently been developed to perform a simultaneous joint analysis of 2D imaging and 1D spectroscopic data.}

\myrev{Inspired by these approaches, we propose a spectrophotometric decomposition technique using images and spectra jointly to effectively separate the AGN and host galaxy components. In this work, we systematically analyze the host galaxy properties of the Dark Energy Spectroscopic Instrument \citep[DESI;][]{DESI2016a.Science, DESI2022.KP1.Instr} quasars at $0.1<z<1$. While previous studies often focused on compact quasars, we investigate quasars with extended morphologies in the optical. The prominent existence of host galaxies in these objects makes them suitable for our joint decomposition pipeline. Utilizing our new method, we derive the physical properties of the host galaxies and their SMBHs.}

The outline of this paper is as follows. In Section \ref{sec:sample}, we introduce the imaging and spectral data that we use, as well as our selection of DESI quasars with extended morphologies. Section \ref{sec:data_analysis} presents the details of our spectrophotometric modeling method. \myrev{In Section \ref{sec:mocktest}, we perform mock tests to evaluate the reliability and robustness of our approach.} We describe the derived properties of the SMBHs and their host galaxies in Section \ref{sec:results}. \myrev{Section \ref{sec:discussion} discusses the physical implications of our results in the context of quasar-host co-evolution.} We summarize our work in Section \ref{sec:summary}. Throughout this paper, we adopt AB magnitudes for brightness representation, and assume a flat $\Lambda$CDM cosmology with $H_0 = 70\, \mathrm{km} \, \mathrm{s}^{-1} \, \mathrm{Mpc}^{-1}$, $\Omega_{\Lambda} = 0.7$, and $\Omega_m = 0.3$.

\section{Data and Quasar Sample} \label{sec:sample}

\begin{figure*}[htbp]
    \centering
    \includegraphics[width=0.7\linewidth]{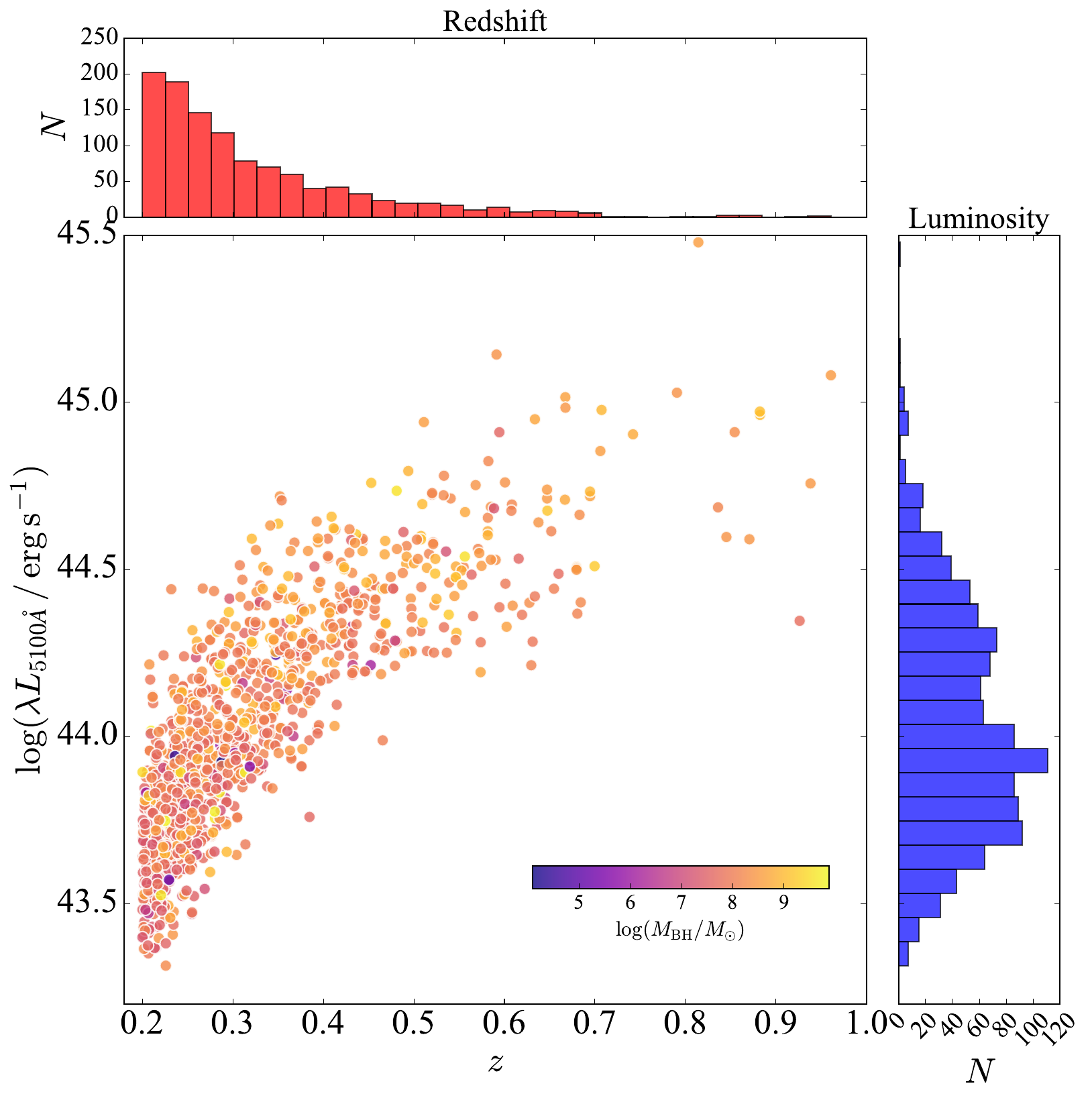}
    \caption{Redshift and luminosity distributions of our sample of DESI quasars with extended morphologies, color-coded by the estimated $M_{\mathrm{BH}}$. The majority of the sample is at $z<0.6$. The luminosity is represented by the monochromatic luminosity at 5100\AA, which is derived from the DESI spectra.}
    \label{fig:sample_distribution}
\end{figure*}

We combine the DESI spectra with the HSC deep imaging data to construct our parent quasar sample. 
The combination of the high-resolution DESI spectra and the high-quality HSC broadband images enables robust component decomposition for quasars. In this work we focus on relatively low-redshift ($0.1<z<1$) quasars with extended morphologies in the HSC images (Figure \ref{fig:sample_distribution}).

\subsection{DESI}

As a robotic, fiber-fed, highly multiplexed spectroscopic surveyor, DESI is among the most powerful and efficient multi-object spectrographs in operation \citep{DESI2022.KP1.Instr}. With 5000 fibers mounted on its focal plane and a field of view of $\sim 3 \arcdeg$, DESI is near the end of completing its 5-year spectral survey that will obtain spectra for around 50 million stars, galaxies, and quasars over approximately 14,000 deg$^{2}$ of the sky, including 3 million quasars \citep{DESI2016a.Science, DESI2016b.Instr, FocalPlane.Silber.2023, Corrector.Miller.2023, FiberSystem.Poppett.2024}. The high spectral resolutions of DESI over the optical wavelength ranges of three channels (blue: 3600 -- 5900 \AA, R $\sim$ 2700; green: 5660 -- 7220 \AA, R $\sim$ 4200; and red: 7470 -- 9800 \AA, R $\sim$ 4600) allow detailed modeling of the AGN host galaxy and the measurement of their physical properties. 
DESI uses the pipeline of \citet{SurveyOps.Schlafly.2023} to tile the survey, to plan and optimize observations along the campaign, and its spectra were reduced with pipeline from \citet{Spectro.Pipeline.Guy.2023}. 

The DESI quasar selection was realized through the Legacy Imaging Survey \citep{DESI_img_DR9, BASS.Zou.2017, Dey_2019} and tested to be robust through validation \citep[e.g.,][]{2023ApJ...944..107C}. 
Several methods were applied for quasar classification by DESI: Redrock, a template fitting code \citep{RedrockQSO.Brodzeller.2023}; \mgii\ Afterburner that uses broad \mgii\ lines \citep{2023ApJ...944..107C}; and QuasarNET, a deep convolutional neural network (CNN) that excels at classification and redshift estimation \citep{quasarnet,Farr_2020}. The reliability of redshift determination and the performance of these methods were verified by visual inspection \citep{2023ApJ...943...68L, Alexander_2023}. We use the published dataset of DESI, called Iron, which includes the Survey Validation (SV) and Year 1 phase survey spectra as part of the First Data Release(DR1, \citealp{2025arXiv250314745D}). It contains spectra for about 1.4 million quasars. 
With DESI DR1, cosmological results from the baryon acoustic oscillations measurements \citep{Cosmo.BAO.KP7} and the full-shape analysis \citep{DESI2024.VII.KP7B} were accomplished.
We select quasars from the value-added catalog (VAC) constructed by the DESI group (SPECTYPE is `QSO'), and crossmatch them with the HSC imaging data. The Signal-to-Noise ratios (S/N) of the spectra are calculated as flux divided by the flux error (converted from \textsc{ivar}), which is $1\sigma$ value. For the following analyses, the spectra are required to have a mean S/N higher than 5 per pixel.

\subsection{HSC}

The imaging data used in this study is from the third data release of HSC-SSP Survey \citep{HSC_DR3_2022}. 
The HSC images in the five broad bands $g,r,i,z,y$ and their variance maps, masks, and PSF models for each DESI quasar are generated using the HSC software pipeline \texttt(hscPipe) v8, which has been customized from the LSST Science Pipelines \citep{hscpipe_2018PASJ...70S...5B, LSST_pipe_2019ASPC..523..521B}.
We crossmatch the HSC PDR3 wide field images with our bright DESI quasars to ensure a large sample size (PDR3 ultra deep field has little overlap with DESI quasars). For each crossmatched quasar, we make a $20 \arcsec \times 20 \arcsec$ HSC cutout data-cube (image, mask, variance) in each band for the following analyses.
We adopt the PSF models provided by the pipeline since they have been well tested \citep[e.g.,][]{HSC_weak_lensing2023PhRvD}.

\subsection{Quasars with extended morphologies}  \label{sec:sample_selection}

\begin{figure*}[htbp]
    \centering
    \includegraphics[width=0.9\linewidth]{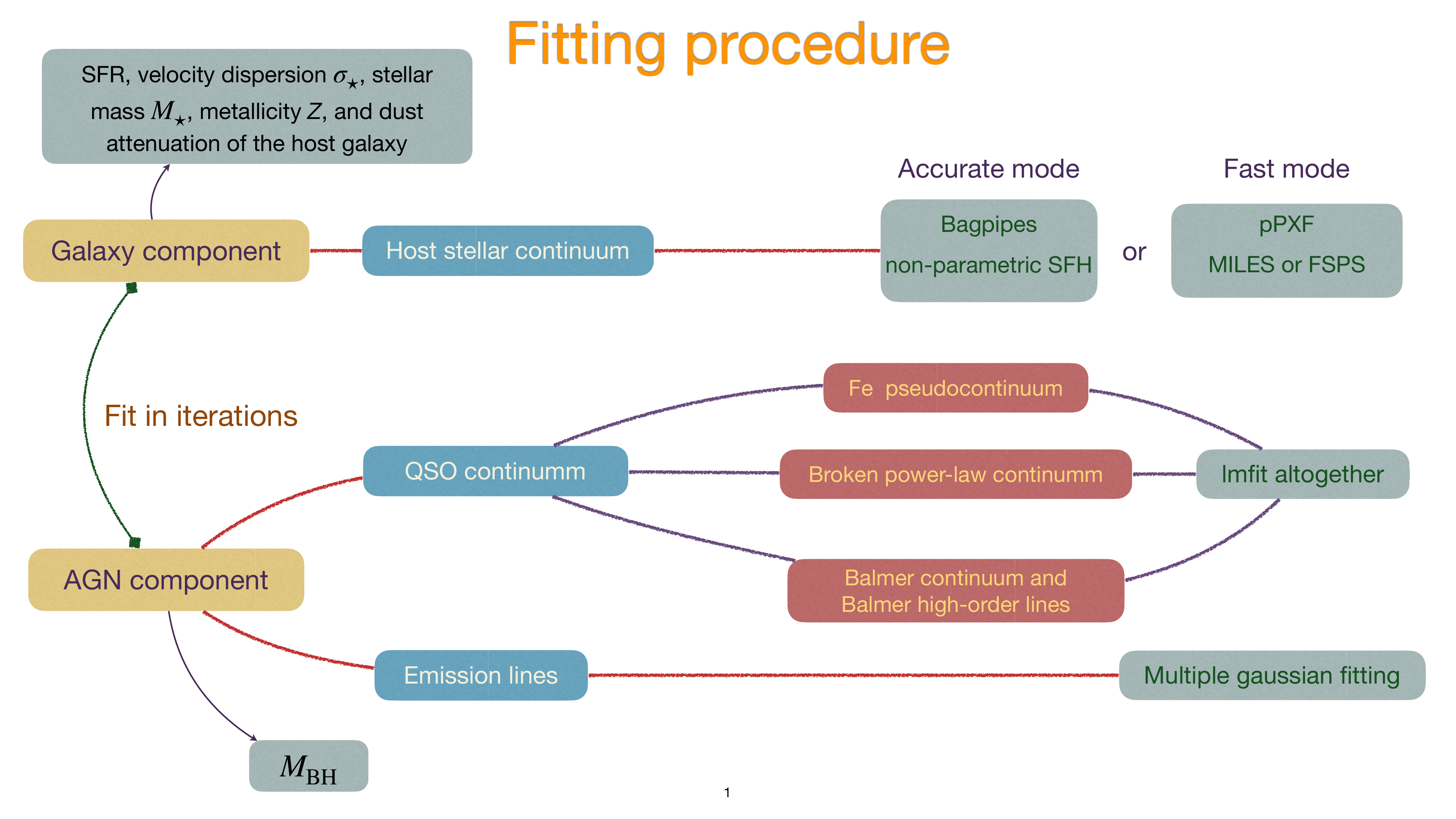}
    \caption{Flowchart explaining how our pipeline works. This is a concise demonstration for the fitting iterations. Details are provided in Section \ref{subsec:spectra decomp}.}
    \label{fig:flow_chart_code}
\end{figure*}

The selection process of our sample of DESI quasars is described below. 
\begin{enumerate}
 \item[$(1)$] From DESI DR1, we select quasars with their DESI Legacy Survey \citep[DR10;][]{DESI_img_DR9, Dey_2019} imaging morphologies classified as `non-psf' and their spectra defined as `QSO' by the DESI RedRock and \mgii\ afterburner classification pipelines. This results in a preliminary sample of DESI quasars with extended morphologies. The sample size is $\sim 163$ k.
 \item[$(2)$] We select the brightest DESI quasars with $\lambda L_{\lambda} (5100\AA) > 10^{43.3} \, \mathrm{erg}\, \mathrm{s}^{-1}$ from the preliminary sample. The selected sample size is $\sim 20$ k.
 \item[$(3)$] We crossmatch the sample with the HSC wide field images, and focus on an intermediate redshift range of $0.1<z<1$. This results in a sample of 3694 DESI quasars.
\end{enumerate}

We briefly explain the above selection procedure. The `non-psf' morphology types for DESI Legacy Survey images include \sersic profiles \textsc{SER}, exponential profiles \textsc{EXP} for spiral galaxies, and deVaucoulers profiles \textsc{DEV} for elliptical galaxies. The DESI quasar spectral classification criterion means that the SPECTYPE of these selected objects is `QSO' in the Iron VAC catalog. We start from the brightest DESI quasars since the spectral S/Ns of faint quasars are very low and important spectral features of their host galaxies cannot be detected. The redshift range of our sample is constrained to be within $0.1<z<1$, which is most suitable for single-epoch black hole mass $M_{\mathrm{BH}}$ measurements and the identification of the host galaxy features.
The lower redshift limit of $z=0.1$ is applied to reduce the impact of host galaxies in the optical images that cannot be well modeled by a single \sersic profile (see Section \ref{subsec:image decomp}). The choice of \sersic profile to depict our host galaxy morphology has also been verified through mock tests. We impose an upper redshift bound at $z=1$ to make sure that important emission and absorption lines are within the observed wavelength ranges of DESI (3600 -- 9800\AA). 

We select 3694 quasars from the above procedure. After further spectrophotometric decomposition and visual inspection of the spectral quality, our final sample size of DESI quasars with prominent host galaxy components is 1126. In Figure \ref{fig:sample_distribution}, we present the luminosity and redshift distributions of the quasars. The majority of these quasars are at $z<0.6$.

\section{Data Analysis} \label{sec:data_analysis}

After crossmatching the DESI spectra with HSC images, we leverage the imaging information to set constraints on the host galaxy to AGN flux ratio. The logic flow of our iterative fitting procedure is shown in Figure \ref{fig:flow_chart_code}. We will present the mock test results in Section \ref{sec:mocktest} to demonstrate the reliability of our spectrophotometric approach. Here we briefly explain our spectrophotometric decomposition method. We first perform an imaging/morphological decomposition using five HSC broad bands ($g, r, i, z, y$) and set constraints on the flux ratios of host galaxy to AGN in different bands. Combining the flux ratios, we perform a global fitting of the  quasar spectra and derive host galaxy properties. This method can help us break the degeneracy of the AGN and host galaxy components in the spectral fitting and reliably determine the host galaxy properties (see the upper left box of Figure \ref{fig:flow_chart_code}).

\subsection{Imaging decomposition}\label{subsec:image decomp}

Imaging data are fitted with \texttt{GalfitM} \citep{2022A&A...664A..92H_galfitm}, and the imaging decomposition is achieved with two iterations of \texttt{GalfitM} fitting \myrev{(see Figure \ref{fig:galfitm_res1}, with an additional example provided in Appendix \ref{sec:appendix_A})}. The host galaxy photometry decomposition results are taken as priors for the galaxy component in the spectral fitting. They are also included as data points in the subsequent \texttt{Bagpipes} \citep{10.1093/mnras/sty2169, 2019MNRAS.490..417C} spectrophotometric joint modeling. The procedure starts with masking of nearby contamination. Mask maps made by \texttt{Photutils} \citep{larry_bradley_2023_7946442_photutils} ensures the removal of all contamination from nearby sources within the HSC $20 \arcsec \times 20 \arcsec$ cutout images. The central 5 pixels of AGNs with very sharp central brightness profiles are also masked. 
After masking, \texttt{GalfitM} multi-band fitting is accomplished with a single \sersic component and a PSF component. The PSF component is fitted with the PSF model  constructed by the HSC weak lensing working group. The free parameters of the \sersic component include a \sersic index, magnitude, effective radius ($R_{e}$), position angle (PA), and axis ratio (AR), and their initial values are provided by \texttt{statmorph} \citep{2019MNRAS.483.4140R_statmorph}.
\myrev{After the first round of the model fitting, we repeat the procedure in the second iteration. To better model the host galaxy contribution, we derive initial values for the \texttt{GalfitM} fitting with the previously fitted central PSF components subtracted from the original images. We utilize \texttt{statmorph}, an image fitting pipeline optimized for galaxy morphology characterization, to provide the initial morphological values. This second iteration significantly improves the host galaxy model.}

\myrev{To break the spectral degeneracy, we utilize these imaging decomposition results to constrain the host galaxy flux within the DESI fiber. Instead of adopting the \texttt{GalfitM} analytical models for the host galaxy contribution, we perform aperture photometry directly on the PSF-subtracted images (i.e., the original observed images with the best-fit AGN PSF removed), utilizing a $1.5^{\prime\prime}$ diameter aperture. Because we extract fluxes from the actual residual images, this procedure inherently preserves the fiber flux.}

\myrev{The uncertainties provided by \texttt{GalfitM} are statistical. We apply a 20\% systematic error floor to the derived host galaxy photometry to account for two primary sources of uncertainty. First, cross-instrument calibration and aperture effects: differences in atmospheric seeing and fiber-positioning offsets between the HSC imaging and DESI spectroscopy introduce uncertainties in the flux determination \citep[e.g.,][]{2016AJ....151....8Y}. Second, template mismatch: applying a $10\%-20\%$ error floor is standard practice in SED fitting to account for inherent uncertainties in stellar population synthesis models \citep[e.g.,][]{2013ARA&A..51..393C} and to prevent high signal-to-noise photometric points from overpowering the spectroscopic constraints \citep[e.g.,][]{2008ApJ...686.1503B, 2020MNRAS.495..905R, 2021MNRAS.505..540T}. During spectral fitting, this photometry serves as a prior, allowing the absolute normalization of the host template to adjust within this 20\% range and break the AGN-host degeneracy.}

\begin{figure*}[htbp]
    \centering
    \includegraphics[width=0.8\linewidth]{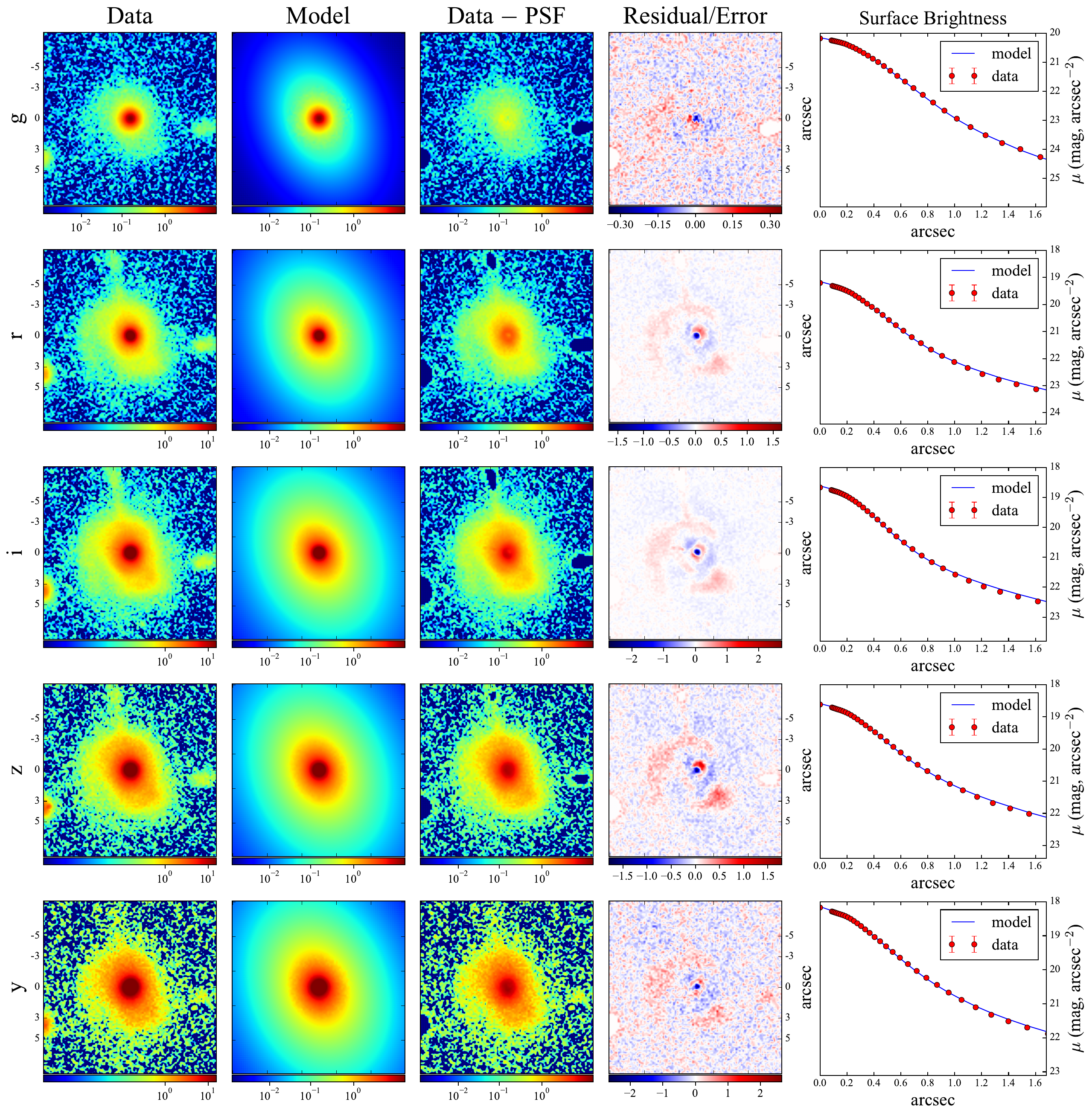}
    \caption{Representative example of the imaging decomposition with \texttt{GalfitM} on a quasar at $z = 0.51$. The images from top to bottom are HSC $g, r, i, z, y$ five broadband images, respectively. The image size is about $8\arcsec$ on a side. The five columns from left to right show the following information: $\left(1\right)$ observed HSC images; $\left(2\right)$ the constructed \texttt{GalfitM} best-fitting models consisting of a point source plus a single \sersic host galaxy, $\left(3\right)$ the images after the subtraction of the best-fit point source model; $\left(4\right)$ The fitting residual; and $\left(5\right)$ the surface brightness profile provided by fitting ellipses centered at the point source. The scales marked on the right side of the 4th column demonstrate the spatial sizes in units of arcsec. The colors in the images reflect the flux level of individual pixels as shown by the color bars. The residual maps suggest that the model fit is robust.}
    \label{fig:galfitm_res1}
\end{figure*}{}

\subsection{Spectral fitting} \label{subsec:spectra decomp}

\begin{figure*}
    \centering
    \includegraphics[width=\textwidth]{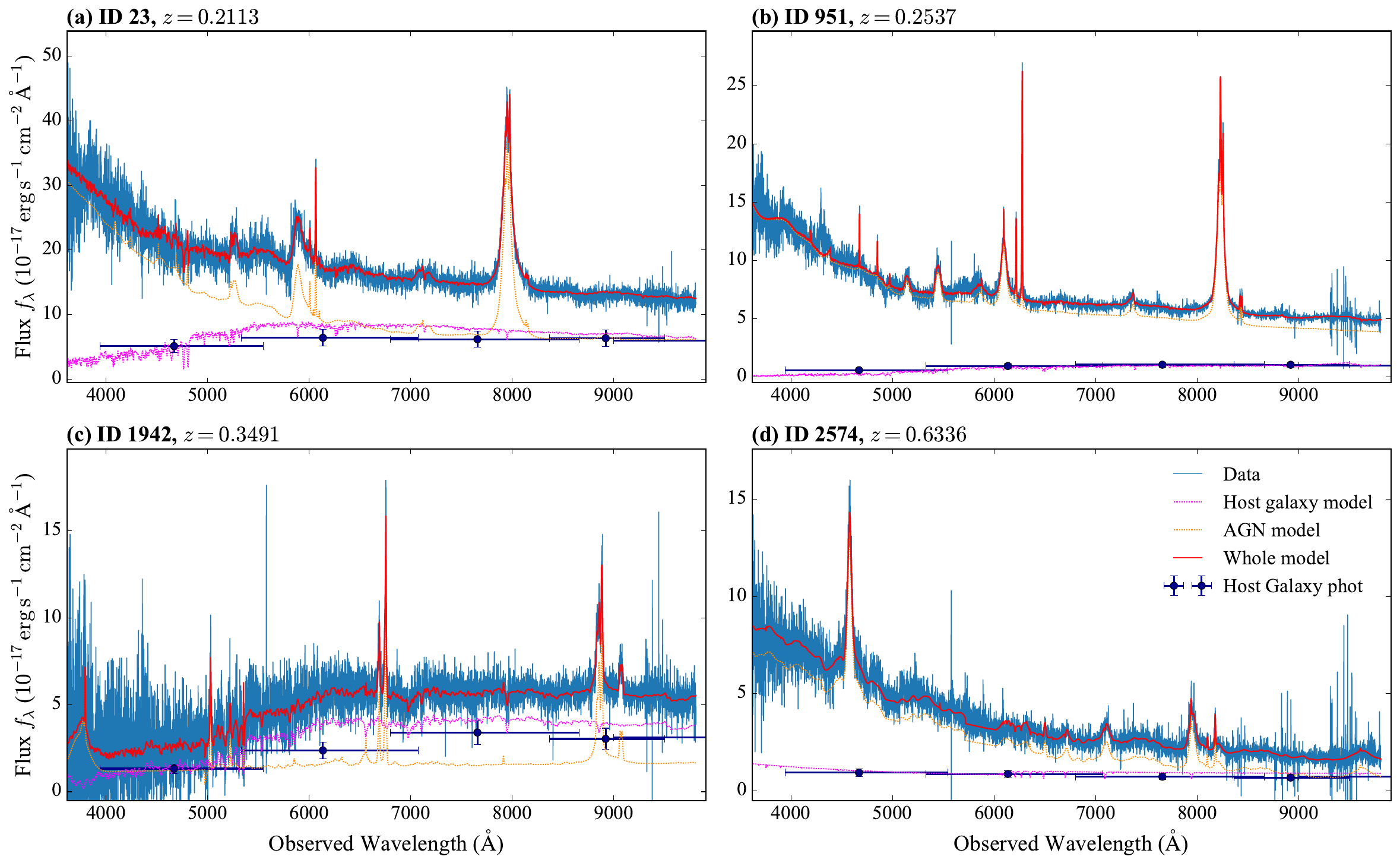}
    \caption{\myrev{Example of spectrophotometric fitting results for four quasar host galaxies. }
    The blue solid line shows the original observed spectrum. 
    The magenta densely dotted and dark orange densely dotted lines indicate the decomposed host galaxy spectrum and AGN model spectrum, respectively. 
    The red solid line indicates the final composite model spectrum. 
    The dark blue dots with error bars show the host galaxy photometry results converted from the GalfitM photometry decomposition. 
    The quasar in panel (a) represents the case where the host galaxy and AGN spectral components are relatively comparable, and the AGN component is bright, dominating the rest-frame UV bands. 
    Panel (b) represents the case where the central AGN is very bright, dominating the whole observed spectrum. 
    Panel (c) represents the case where the host galaxy dominates the whole optical wavelengths, and the central AGN is faint. 
    Panel (d) represents the case where the object is located at a relatively high redshift, and the AGN component is bright, dominating the rest-frame UV bands. It also produces prominent Mg\,{\sc ii} emission lines.
    }
    \label{fig:four examples of example spectral fitting} 
\end{figure*}

We combine the \texttt{PyQSOFit} \citep{2018ascl.soft09008G}, \texttt{pPXF} \citep{Cappellari2023}, and \texttt{Bagpipes} fitting codes in the spectral fitting procedure (see Figure \ref{fig:four examples of example spectral fitting} for an example). We discuss in details below the iterative fitting procedure shown in Figure \ref{fig:flow_chart_code}. The model of the quasar spectral component includes two components, the quasar continuum and emission lines. We attribute all emission lines to the quasar component, while all subtle absorption lines are accounted for in the stellar continuum component of the host galaxy. 
We use two modes of spectral modeling for host galaxies, referred to as `accurate mode' and `fast mode'. In the `accurate mode', the host galaxy photometry data points and spectra are modeled with \texttt{Bagpipes} through the non-parametric SFH method, which is computationally expensive but gives more reliable estimation of physical parameters. In the `fast mode', we use \texttt{pPXF} to replace \texttt{Bagpipes} to save computing time.

The quasar continuum comprises three parts, including a Fe pseudo-continuum, a broken power-law continuum originating from the accretion disk of the central black hole, and a Balmer continuum.
The Fe templates in the UV band are adapted from Vestergaard01 \citep{2001ApJS..134....1V}, and Tsuzuki06 \citep{Tsuzuki_2006}, and the Fe templates in the optical bands are the results from \citet{1992ApJS...80..109B}. The Balmer high-order lines forming pseudo-continuum redward of the Balmer edge and Balmer continuum blueward of the Balmer edge are modeled following the method described in \citet{Hu_2008}. The broken power-law continuum is modeled with a double power-law function. The above components are fitted with \texttt{lmfit} python package in selected continuum fitting windows. 

Emission lines in the quasar spectra are fitted with several Gaussian functions. The criterion separating the broad and narrow components of the emission lines is set to be $1500$ \kmps. The typical emission lines accounted for in our model include Balmer lines from \ha\ to \hdelta, \mgii, \civ, \oiii\, and \siii. \oii\ and \oiii\ are commonly known SFR indicators for galaxies. The emission lines contain rich information, and they are modeled in a similar way in \texttt{PyQSOFit}. 

The host stellar continuum is fitted with the non-parametric SFH method implemented in \texttt{Bagpipes}, assuming 14 age bins covering the timescale from the cosmic Big Bang to the current observed time of the object. The non-parametric SFH adopts a student-$t$ prior, as in \texttt{Prospector$- \alpha$} \citep{Leja_2019}. 
\myrev{The velocity dispersion $\sigma_{\star}$, stellar mass $M_{\star}$, metallicity $Z$, dust attenuation, and SFR of the host galaxy are all globally fitted and derived by \texttt{Bagpipes}. SFRs are from the \texttt{Bagpipes} modeling results rather than empirical estimations from narrow emission lines (e.g., \oii). The majority of our extended quasar hosts are quiescent or post-starburst galaxies, and thus do not have prominent \oii\ emission.}

The fitting procedure starts with the decomposed host galaxy photometry. \texttt{Bagpipes} is utilized to accomplish the SED fitting of the decomposed five-band host galaxy photometry. Through this first step, the \texttt{Bagpipes} fitted host galaxy model spectrum across the whole wavelength range (from infrared to far UV, interpolated into the same observed wavelength range as the same resolution of the input observed spectra) is used as a rough estimate of the host galaxy spectrum. Then this evaluated host galaxy component is subtracted from the original spectrum of the quasar, and the residual spectrum is considered as the pure quasar component. After the construction of the model quasar spectrum to fit the disentangled pure quasar component, this model spectrum is also subtracted from the original spectrum, and the residual spectrum is taken as the pure host galaxy spectrum, which is combined with the decomposed host galaxy photometry to be fitted by \texttt{Bagpipes} simultaneously. After the spectrophotometric \texttt{Bagpipes} fitting of the host galaxy component, the constructed model host galaxy spectrum is again subtracted from the original observed spectrum, taking the residual as the pure quasar component to be fitted. The fitting procedure then iterates.
Every iteration of our composite fitting begins with the fitting of the deemed pure quasar component and finishes with the subsequent host galaxy component fitting with \texttt{Bagpipes}. 
The composite fitting iterations stop when the modeling of the two major components converges, that is, when the fitted physical parameters of the quasar and host galaxy component do not deviate too much from the last iteration, namely less than $5\%$. 

Since the \texttt{Bagpipes} fitting of the host galaxy spectral component is computationally time-consuming, our code also provides an alternative method in the iterations of global composite fitting where we can replace the role of \texttt{Bagpipes} by \texttt{pPXF}. The \texttt{pPXF} fitting of the host galaxy stellar continuum is mainly achieved by constructing the models using the \texttt{MILES} and Flexible Stellar Population Synthesis (\texttt{FSPS}) stellar continuum templates.

\begin{figure*}
    \centering
    \begin{minipage}{0.485\textwidth}
        \centering
        \includegraphics[width=\textwidth]{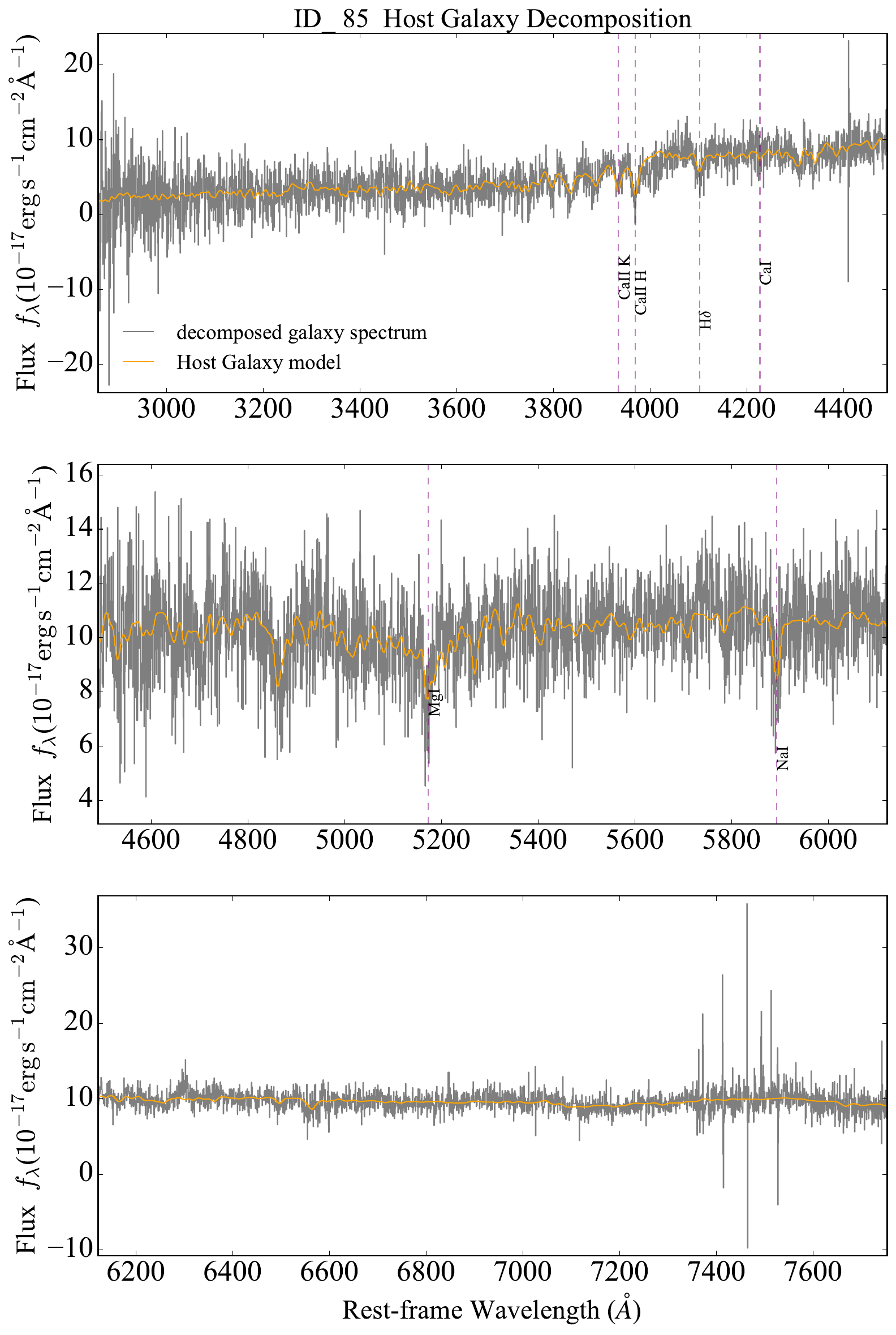}
    \end{minipage}%
    \hfill %
    \begin{minipage}{0.497\textwidth}
        \centering
        \includegraphics[width=\textwidth]{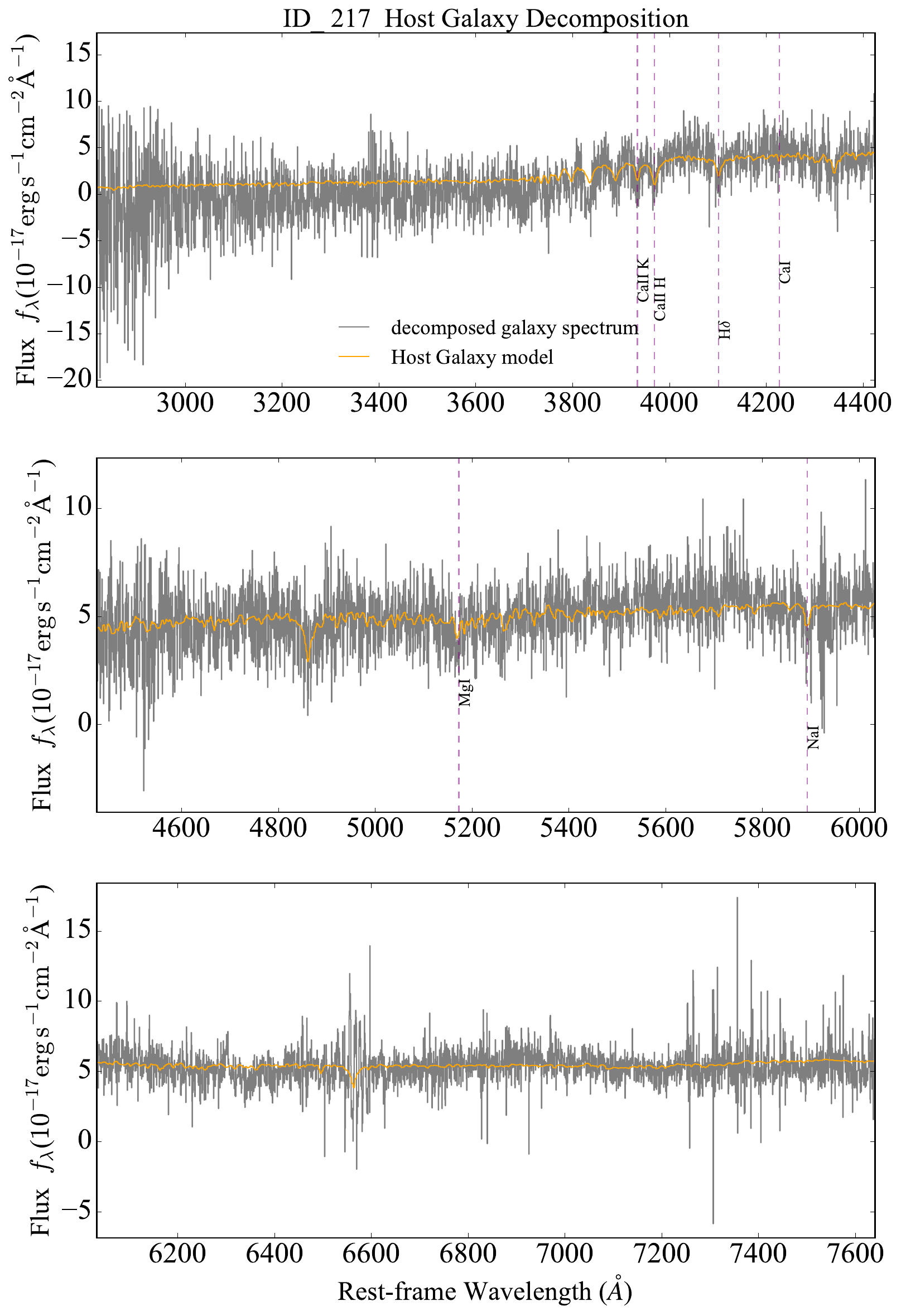}
    \end{minipage}
    \vspace{0.5em} %
    \caption{Demonstration of the decomposed host galaxy spectra. The grey lines represent our final spectral results of the spectrophotometric decomposition method and the yellow lines represent the corresponding \texttt{Bagpipes} galaxy models. We highlight the important absorption line features with the purple dashed lines. The host galaxies of the two quasars show prominent post-starburst features.}
    \label{fig:decomposed_host_galaxies_in_detail}
\end{figure*}

\subsection{Host galaxy stellar velocity dispersion}

The velocity dispersion measurements here are based on the \texttt{Bagpipes} fitting of absorption lines. Due to the redshift distribution of our sample, we mainly capture the \mgi\ and \caii\ HK absorption lines to characterize the host stellar velocity dispersion $\sigma_{\star}$, as shown in Figure \ref{fig:decomposed_host_galaxies_in_detail}.
We correct the instrument broadening effect in the final host $\sigma_{\star}$ measurements.

The uncertainties and advantages of the choice of the method are discussed below. We set an upper limit of the host galaxy velocity dispersion to be $\sigma_{\star} \leq 1000$ \kmps, which is reasonable within uncertainties. The \texttt{Bagpipes} fitting results show that some quasars have very large velocity dispersions that approach this limit due to the low S/N of the decomposed host galaxy spectra. These quasars with such unphysical measurements are typically AGN dominated sources with very weak host galaxy signals. We thus exclude these quasars from our $M_{\mathrm{BH}} - \sigma_{\star}$ relation in Section \ref{subsec:M-sigma relation}.

The host $\sigma_{\star}$ values are measured from the global fitting of the decomposed host galaxy spectra. We fit a broadening function convolved with the templates generated by stellar population synthesis to the observed galaxy spectra. We choose to use a global fitting rather than the FWHMs of individual stellar absorption lines due to the following reasons.
\begin{enumerate}
 \item[$(1)$] First, due to the low S/N of the absorption lines in most spectra, the measurement uncertainties are large. Although DESI has high spectral resolutions and can efficiently discriminate between different lines across the observed UV-optical wavelength range, the spectra are usually not good enough to resolve the detailed structures of the stellar absorption lines.  
 \item[$(2)$] Second, due to the different origins of broadening effects, the host $\sigma_{\star}$ results derived from single absorption lines often deviate from their dynamical nature. Global fitting results help reduce this effect. With all the absorption lines taken into account, we can minimize the broadening effects caused by different sightlines, viewing angles, and rotational broadening of the absorption lines, etc.
 \item[$(3)$] \myrev{Third, alternative approaches to circumvent AGN continuum dilution, such as extracting stellar velocity dispersions exclusively from specific narrow stellar absorption features \citep[e.g., the Ca II triplet or Mg I $b$ lines;][]{2006ApJ...641..117G}, are difficult to implement robustly for our current sample. These feature-specific techniques typically demand significantly higher signal-to-noise ratios and spectral resolutions to isolate the narrow absorption lines from the underlying noise and quasar emission. Given the current data quality of our extended DESI quasar spectra, our global spectral fitting approach remains the most optimal and reliable strategy.}
\end{enumerate}

\myrev{Having established the fitting procedure, we verify its robustness against AGN contamination before applying it to the real DESI quasar sample. This validation is detailed in the following section.}

\section{Mock tests and validation}\label{sec:mocktest}

\begin{figure}
    \centering
    \includegraphics[width=0.99\linewidth]{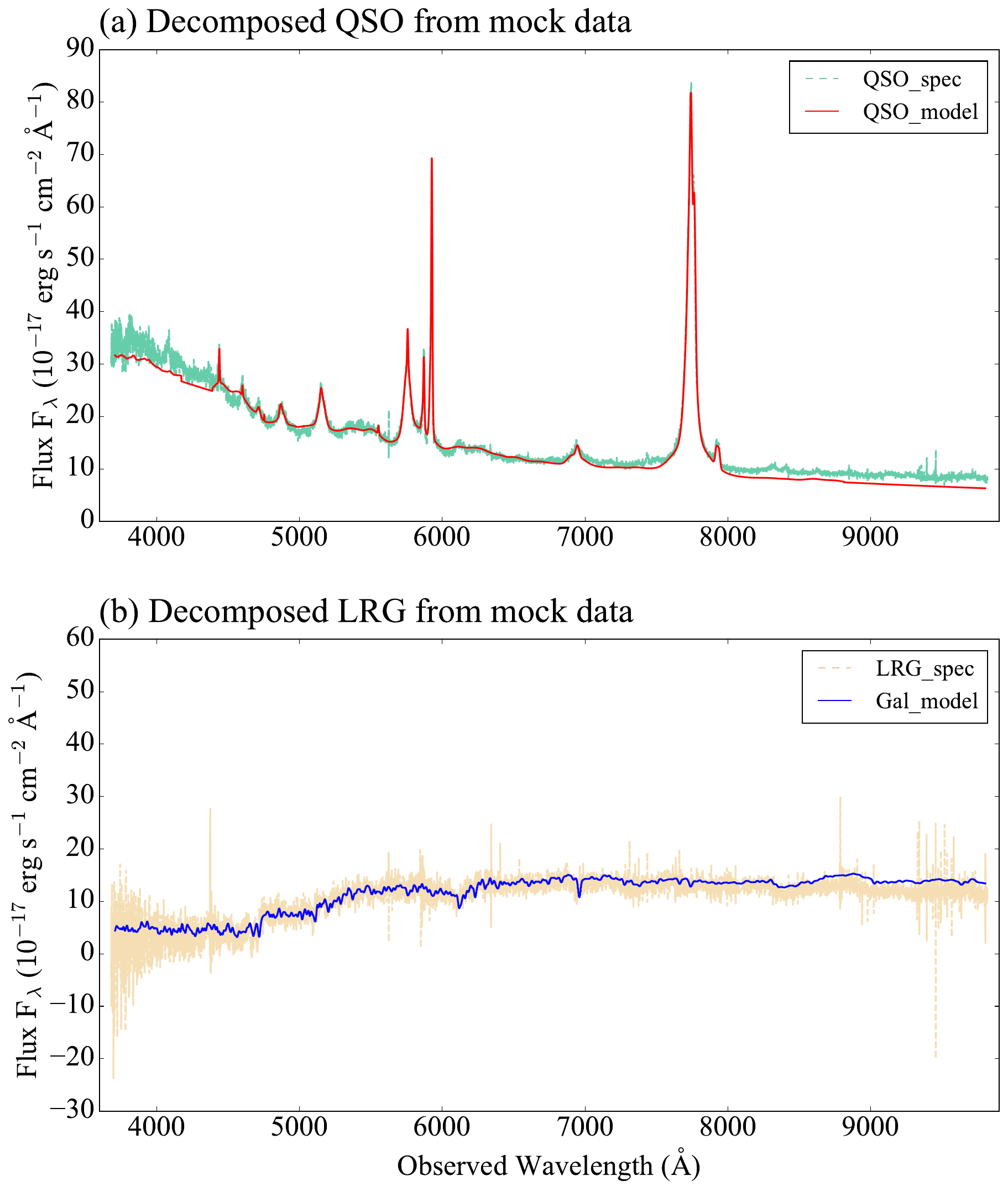}
    \caption{Example of the decomposed spectral components for an AGN and its host galaxy compared with the input spectra. In the upper panel, we plot the decomposed AGN spectrum in red and the input intrinsic AGN spectrum in green. In the lower panel, we show the input LRG spectrum in yellow and the decomposed host galaxy component in blue. In both panels, the input spectra and output results are well consistent.}
    \label{fig:mock test showcase}
\end{figure}

We construct two sets of mock quasars to demonstrate the reliability of our spectrophotometric decomposition pipeline. To construct the first, we combine the brightest quasars and luminous red galaxies (LRGs) from DESI Iron Data Release to mimic the AGNs and host galaxies of DESI quasars. Here we assume that the brightest quasars are completely dominated by their central AGNs. By varying the flux ratios between quasars and LRGs, we adjust the relative strengths of AGN and host galaxy components. An important caveat is that the redshifts of the quasars and the LRGs should be similar ($\Delta z < 0.02$). In Figure \ref{fig:mock test showcase}, we show an example that we can successfully recover the input spectra of a quasar.

\begin{figure*}
    \centering
    \includegraphics[width=0.97\linewidth]{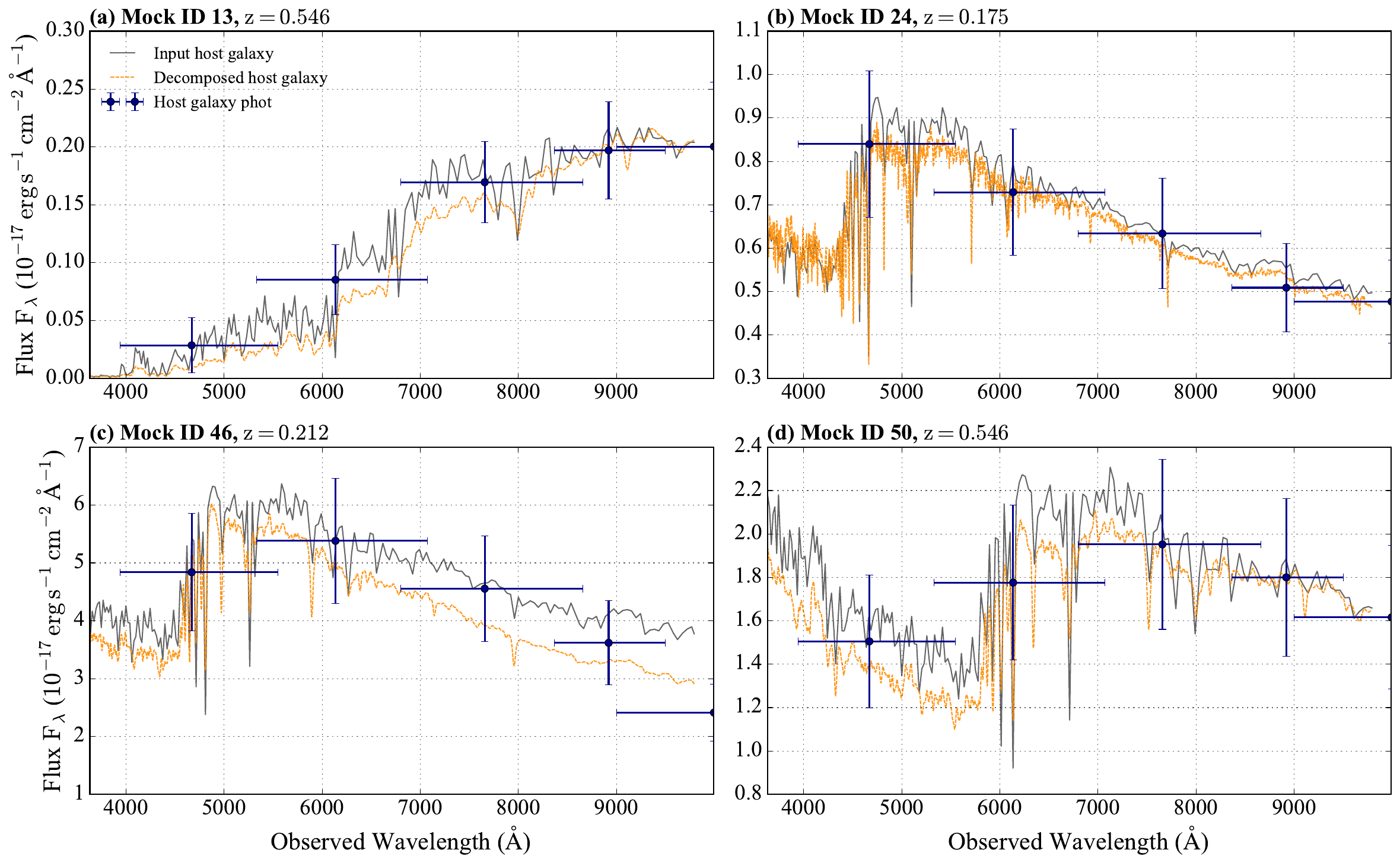}
    \caption{Recovery of the embedded host galaxy spectra in the second mock data set using our spectrophotometric decomposition pipeline. Input spectra combine simulated pure galaxy observation and quasar templates. Here we show four examples which are located at different redshifts with various SFH and stellar mass $M_{\star}$.
    All four panels show consistent decomposed host galaxy components with their input spectra, although the velocity dispersions are not very well modeled for all cases, identical to the results of very large scatter for the measurement of host velocity dispersions in Figure \ref{fig:M-sigma}.}
    \label{fig:recovery-examples}
\end{figure*}

To construct the second set of mock quasars, we use the same bright quasars in DESI as pure AGN templates, while the host galaxy component is generated with \texttt{Bagpipes} and \texttt{Galsim} \citep{2015A&C....10..121R}. We use \texttt{Bagpipes} to generate galaxy spectra with various SFHs, and use \texttt{Galsim} to produce galaxy images with different \sersic profiles. We normalize the mock galaxy images so that their aperture photometry is consistent with the adopted galaxy spectra. These mock quasars cover the full parameter space that we consider.

Below is the detailed description for the mock construction. We construct the SFHs of the host galaxies with 500 different ages. The galaxy metallicities are uniformly distributed between 0.5 and 3. We adopt a Calzetti dust attenuation law with a fixed $A_V = 0.2$. The velocity dispersion is fixed as 200\kmps. By varying the input stellar mass $\log(M_{\star}/M_{\odot})$, we mimic different host galaxy fractions in quasar spectra. The corresponding galaxy images are constructed using \texttt{Galsim}, assuming a single \sersic surface brightness profile. We assign a series of values for the four parameters of galaxy morphology, \sersic index $n$, effective radius $R_{\mathrm{eff}}$, axis ratio $q$, and position angle PA. The \sersic indices $n$ and effective radius $R_{\mathrm{eff}}$ across the five broadband images are produced according to the third-order polynomial function of the effective wavelengths of the optical bands $g,r,i,z,y$. The \sersic indices are required to be in a range of $[0.3, 6]$. The axis ratio $q$ follows a uniform distribution in a ranging of $[0.1, 0.9]$, and the position angle PA follows a uniform distribution in a range of $[0, 90]$ degrees. We adopt PSFs from the HSC official pipeline. 

We then apply our spectrophotometric decomposition technique to the mock quasar spectra. Several examples of the decomposed spectra compared with the input \texttt{Bagpipes} generated ones are depicted in Figure \ref{fig:recovery-examples}. The flux levels of the spectrophotometric decomposition results match well with the intrinsic input data.

Statistical analysis for the decomposition results of the second mock dataset is displayed in Appendix \ref{sec:appendix_A} (Figure \ref{fig:flux level recovery}), which shows a tight relation between the input and output flux values of mock host galaxies. For host galaxies with flux larger than $10^{-16} \, \erg \cdot \rm{s}^{-1} \cdot \mathrm{cm}^{-2} \cdot \text{\AA}^{-1} $ (mainly orange data points), their recovered median fluxes show a one-to-one relation with their input values. This demonstrates that our spectrophotometric approach can successfully recover host galaxy properties if their flux is non-negligible. 
When the flux is very low, there is a large scatter shown in the faint end of the input galaxies. This is mainly due to the large uncertainties introduced by low S/Ns.

\begin{figure*}[htbp]
    \centering
    \includegraphics[width=0.8\linewidth]{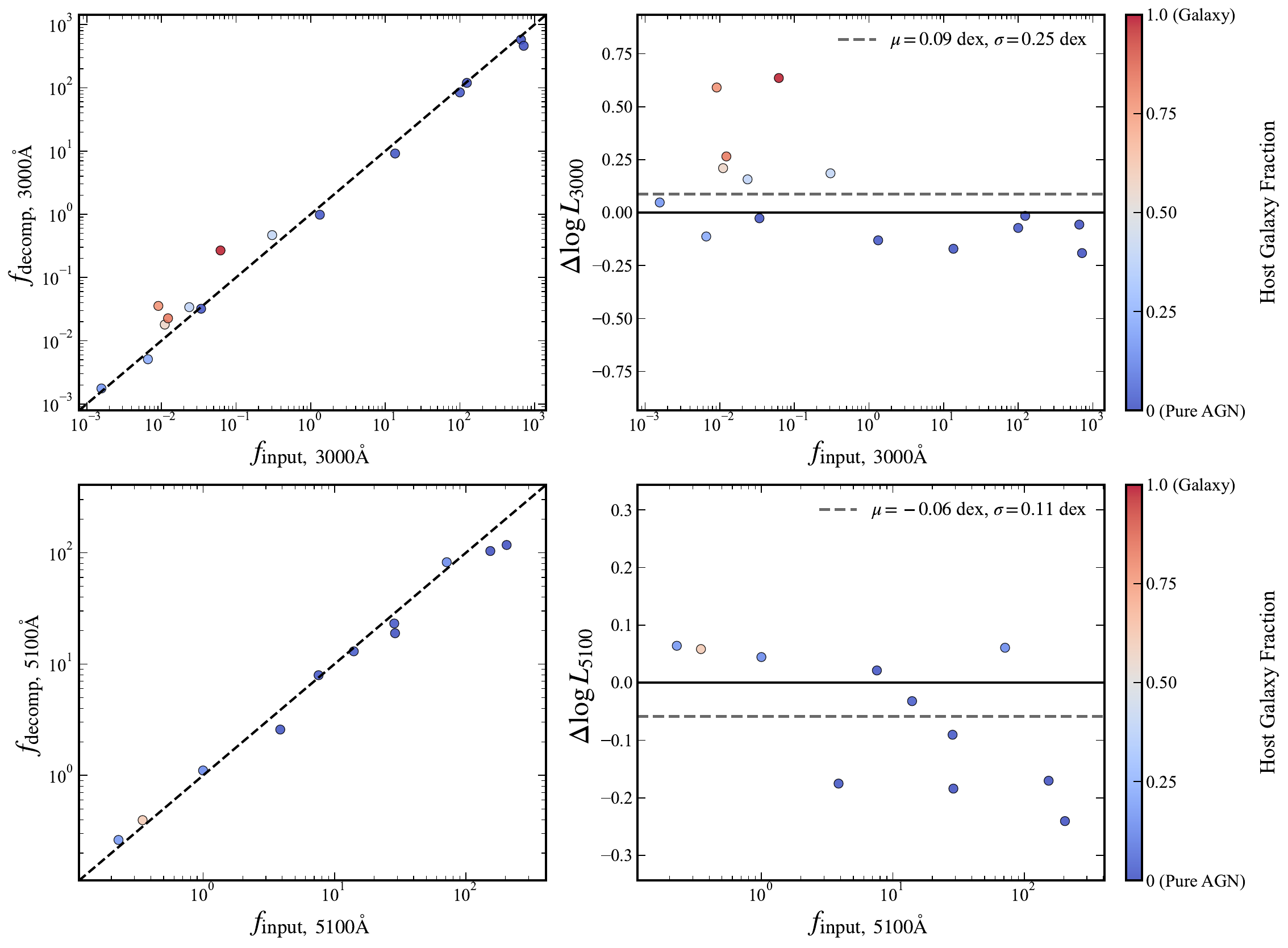}
    \caption{\myrev{Validation of the recovered AGN monochromatic fluxes at $3000$ \text{\AA} (top row) and $5100$ \text{\AA} (bottom row) using our mock dataset. \textit{Left panels:} Comparison between the input and recovered AGN fluxes. The dashed lines represent the one-to-one relations. \textit{Right panels:} Logarithmic residuals, $\Delta \log L$ (in dex), plotted as a function of the input fluxes. The mean ($\mu$) and standard deviation ($\sigma$) of the residuals are indicated by horizontal dashed lines. Data points in all panels are color-coded by the input host galaxy fractions ($f_{\mathrm{host}}$), defined as the ratio of host-to-total flux at rest-frame $5100$ \text{\AA}. All fluxes $f$ are in units of $10^{-17}\,\mathrm{erg\,s^{-1}\,cm^{-2}\,\AA^{-1}}$.}}
    \label{fig:mock_monochromatic_flux}
\end{figure*}

\begin{figure}[htbp]
    \centering
    \includegraphics[width=0.99\linewidth]{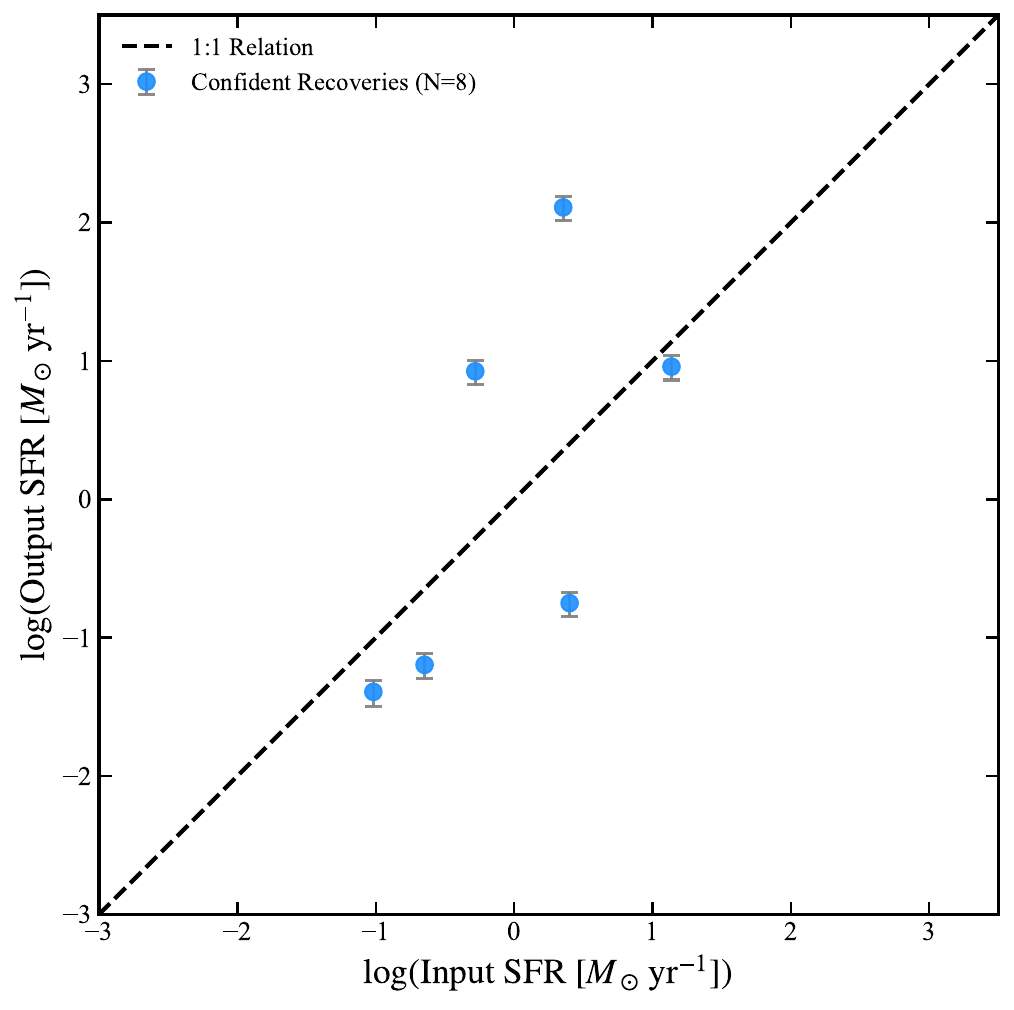}
    \caption{\myrev{Comparison of the input host galaxy SFRs against the recovered SFRs from our mock dataset. The dashed line represents the one-to-one relation. The recovered SFRs are generally consistent with the input values across the sample.}}
    \label{fig:mock_sfr_comparison}
\end{figure}


\begin{figure}[htbp]
    \centering
    \includegraphics[width=\linewidth]{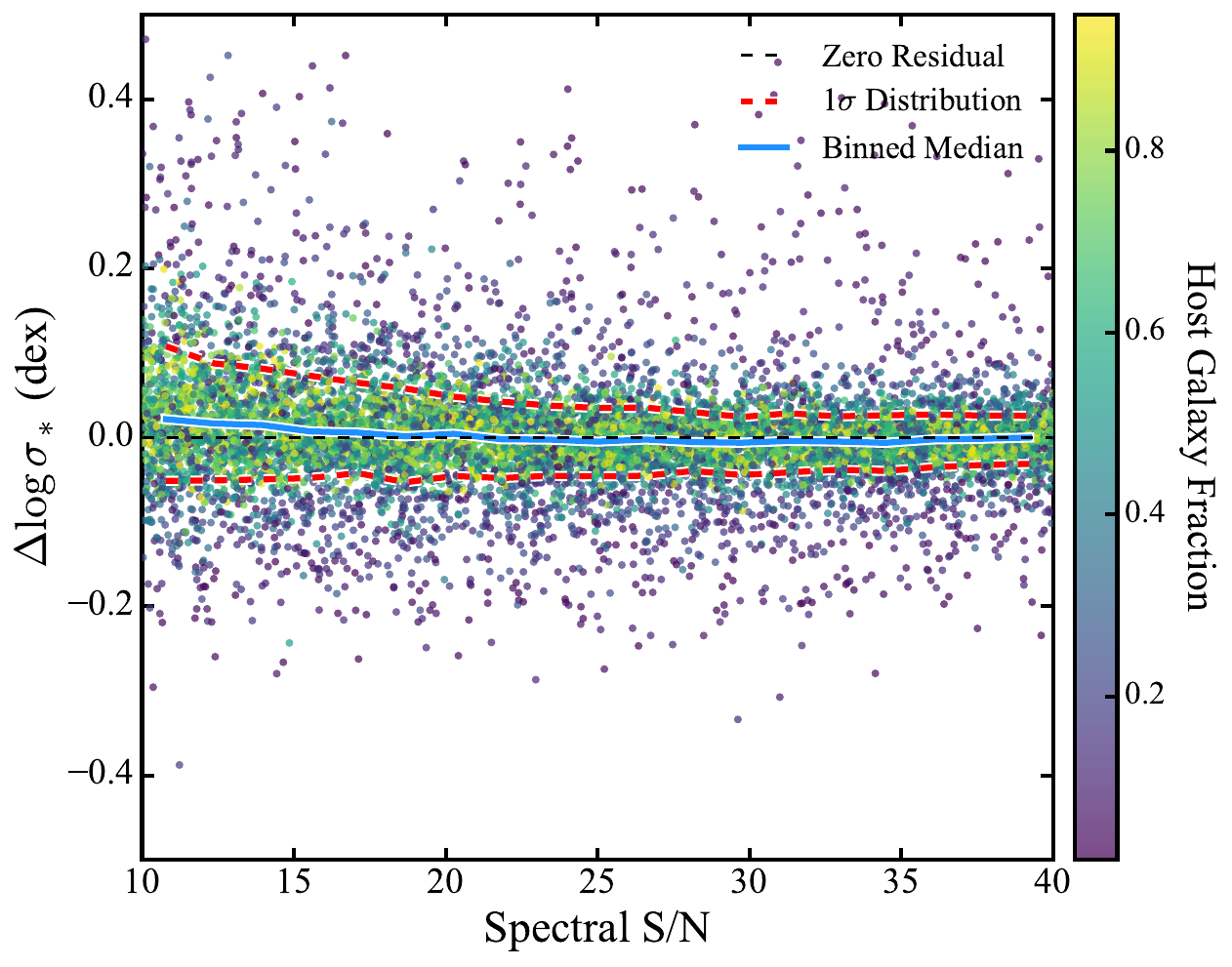}
    \caption{Logarithmic residuals ($\Delta \log \sigma_{\star}$) of the recovered host galaxy stellar velocity dispersion as a function of the input spectral S/N. Data points are color-coded by the host galaxy flux fraction. The binned median of the recovered distribution is shown as a solid blue curve, while the red dashed lines denote the $1\sigma$ scatter envelope. The dashed black line indicates zero residual. The plot demonstrates that the measurement scatter is primarily driven by data quality and the relative host contribution.}
    \label{fig:mock_vd_comparison}
\end{figure}

\myrev{To evaluate the performance of our spectrophotometric decomposition pipeline, we generated a subsample of mock spectra and their corresponding images. This targeted mock dataset was designed to test the reliability of three critical physical parameters extracted by our pipeline: the AGN continuum luminosity, the host galaxy star formation rate (SFR), and the stellar velocity dispersion ($\sigma_{\star}$).}

\myrev{First, regarding the AGN continuum luminosity, which is crucial for black hole mass estimation, we compared the input intrinsic AGN monochromatic fluxes at 3000\AA\ and 5100\AA\ with the output values derived from our pipeline. As illustrated in Figure \ref{fig:mock_monochromatic_flux}, the decomposed fluxes exhibit a tight, one-to-one consistency with the intrinsic input values across a wide range of luminosities. Furthermore, we examined the logarithmic residuals ($\Delta \log L$) as a function of the spectral S/N, redshift, and the input host galaxy fraction ($f_{\mathrm{host}} = F_{\mathrm{host}} / (F_{\mathrm{AGN}} + F_{\mathrm{host}})$ at rest-frame $5100$ \text{\AA}). We found no noticeable trends or systematic deviations in these distributions. This lack of correlation confirms that our pipeline provides unbiased measurements of the AGN continuum luminosity, irrespective of how dominant the host galaxy is.}

\myrev{Second, to demonstrate the reliability of the derived host galaxy SFRs, we compared the input SFRs of our mock galaxies with the recovered values from the decomposition. As shown in Figure \ref{fig:mock_sfr_comparison}, the recovered SFRs align well with the intrinsic values without significant systematic offsets, confirming that the degeneracy between the AGN and host galaxy spectra does not severely bias our SFR estimates.}

Finally, we evaluated the robustness of the stellar velocity dispersion ($\sigma_{\star}$) measurements. This is challenging because the host galaxy stellar absorption features are heavily diluted by the bright AGN continuum and spectral noise. 
Figure \ref{fig:mock_vd_comparison} presents the logarithmic residuals ($\Delta \log \sigma_{\star}$) of the recovered velocity dispersions as a function of the spectral S/N. While the individual measurements exhibit a large scatter, this scatter clearly narrows towards higher S/N and larger host galaxy fractions. In addition, the S/N-binned median consistently aligns with the zero-residual line. These trends indicate that the recovered $\sigma_{\star}$ is statistically unbiased, and the measurement uncertainties are primarily governed by data quality (S/N and AGN dilution) rather than systematic biases in the pipeline.
This large, S/N-driven uncertainty also justifies our choice of employing a global spectral fitting approach rather than fitting specific narrow absorption features (e.g., \caii\ H\&K). Because individual absorption lines are easily overwhelmed by the AGN continuum, localized fitting is often degenerate. Global fitting integrates over all available absorption features and the broadband stellar continuum simultaneously. This effectively maximizes the signal-to-noise ratio and mitigates the impact of AGN and noise dilution.

\section{Results} \label{sec:results}


\myrev{With the reliability of our spectrophotometric pipeline demonstrated through the mock tests in Section \ref{sec:mocktest}, we now present and analyze the decomposition results for our observed quasar sample. As our methodology models the central PSF as the pure AGN emission and attributes the spatially extended component to the host galaxy, the pipeline requires the host galaxy to be sufficiently bright in the HSC imaging. For marginally detected hosts, the derived stellar mass ($M_{\star}$) and other physical properties would be highly uncertain or suffer from systematic biases. Therefore, as detailed in Section \ref{sec:sample_selection}, our analysis focuses on quasars with optically prominent, extended host galaxies. This morphological pre-selection ensures that the DESI quasars in our sample exhibit evident host galaxy components in both the optical imaging and spectral data, making them suitable for robust spectrophotometric decomposition.}

\subsection{SMBH masses}

We use the virial approach to derive the central SMBH masses $M_{\mathrm{BH}}$ of the quasars from their decomposed AGN components. Assuming that the motion of the material in the board line region (BLR) is dominated by the SMBH gravity, the SMBH mass is calculated via
\begin{equation}\label{virial_MBH}
M=f\frac{R_{\mathrm{BLR}} (\Delta V)^2}{G},
\end{equation}
where $f$ is the virial factor depending on the geometry, kinematics, and inclination of the BLR, $R_{\mathrm{BLR}}$ is the size of BLR, $\Delta V$ is the velocity width of the broad emission lines,
and G is the gravitational constant.
Reverberation mapping (RM) provides us with a way of estimating the size of BLR through the continuum flux by the $R-L$ relation \citep[e.g.,][]{peterson1993_RM}. Here we adopt the calibration from \citet{2015ApJ...809..123H_ho_hbeta_SE_mbh},
\begin{equation}
\begin{split}
\label{MBH_adopted}
\log M_{\mathrm{BH}} = \log \left[ \left(\frac{\mathrm{FWHM}_{\mathrm{H}\beta}}{1000 \kmps} \right)^2 \left(\frac{\lambda L_{\lambda} (5100\text{\AA})}{10^{44} \, \mathrm{erg}\, \mathrm{s}^{-1}} \right)^{0.533} \right] \\
+a,
\end{split}
\end{equation} 
where $a=6.91 \pm 0.02$ is applied since the bulge types are not clearly separated in our sample.

\begin{figure}[htbp]
    \centering
    \includegraphics[width=0.99\linewidth]{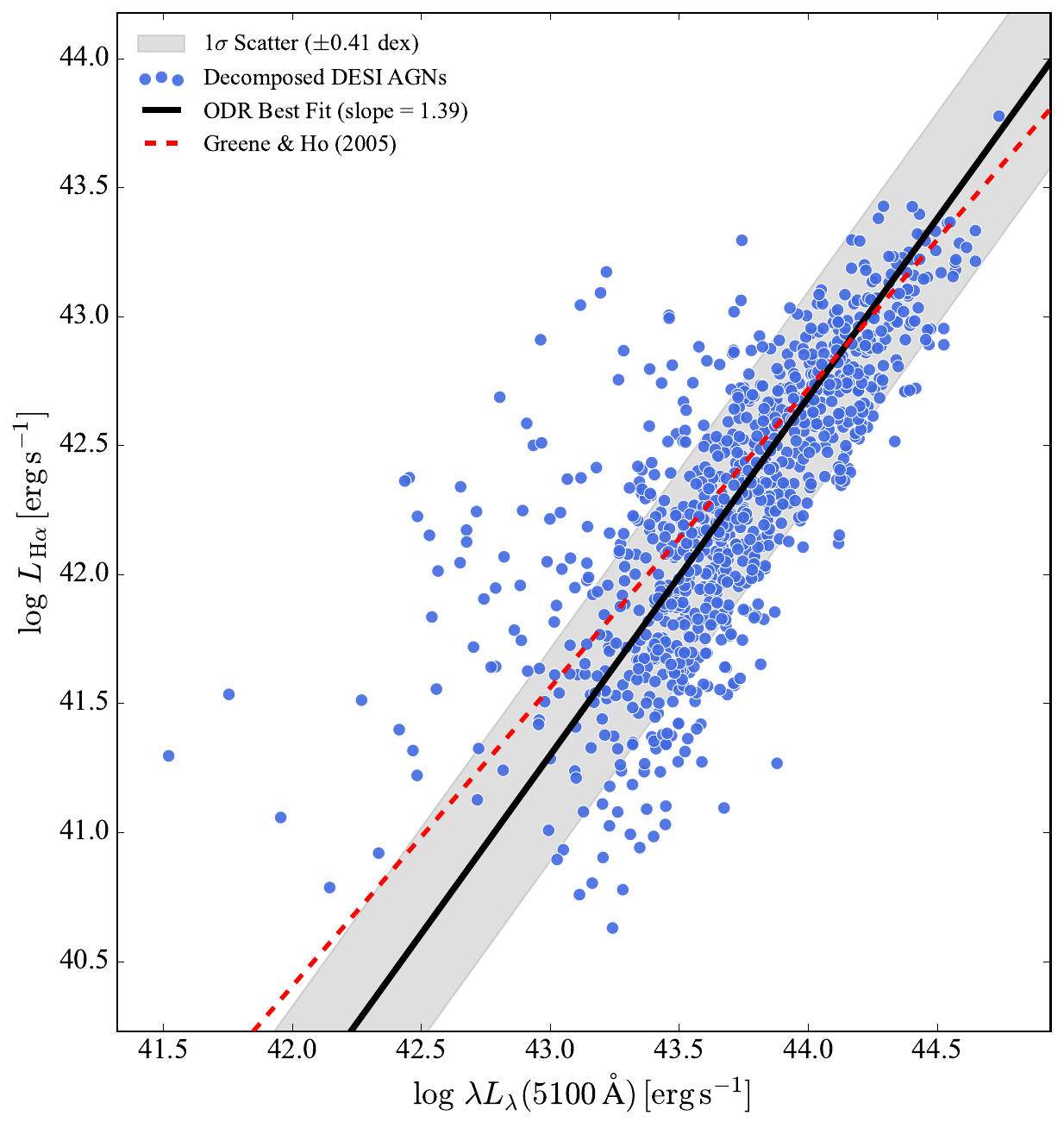}
    \caption{\myrev{Comparison between the decomposed AGN continuum luminosity at $5100$ \text{\AA} ($L_{5100}$) and the broad \ha\ emission line luminosity ($L_{\mathrm{H}\alpha}$). Blue points represent our measurements. The red dashed line denotes the canonical relation for unobscured quasars \citep[e.g.,][]{2005ApJ...630..122G, 2023ApJ...953..142C}. The black solid line shows the orthogonal distance regression (ODR) best fit (slope $= 1.39$), with the grey shaded region indicating the $1\sigma$ scatter ($\pm 0.41$ dex).}}
    \label{fig:LHa_L5100_relation}
\end{figure}

\myrev{To verify that our decomposed AGN continuum luminosities are not biased by systematic over- or under-subtraction of the host galaxy starlight, we test the relationship between the broad \ha\ emission line luminosity ($L_{\mathrm{H}\alpha}$) and the decomposed AGN continuum luminosity at 5100\AA\ ($L_{5100}$). 
As shown in Figure \ref{fig:LHa_L5100_relation}, our sample follows the well-established empirical scaling relation between $L_{\mathrm{H}\alpha}$ and $L_{5100}$ for typical unobscured quasars \citep[e.g.,][]{2005ApJ...630..122G, 2023ApJ...953..142C}. The tight consistency indicates that our spectrophotometric decomposition pipeline reliably isolates the intrinsic AGN continuum.}

\subsection{Properties of host galaxies}\label{sec:host_gal_prop}

\begin{figure}
    \centering
    \includegraphics[width=0.99\linewidth]{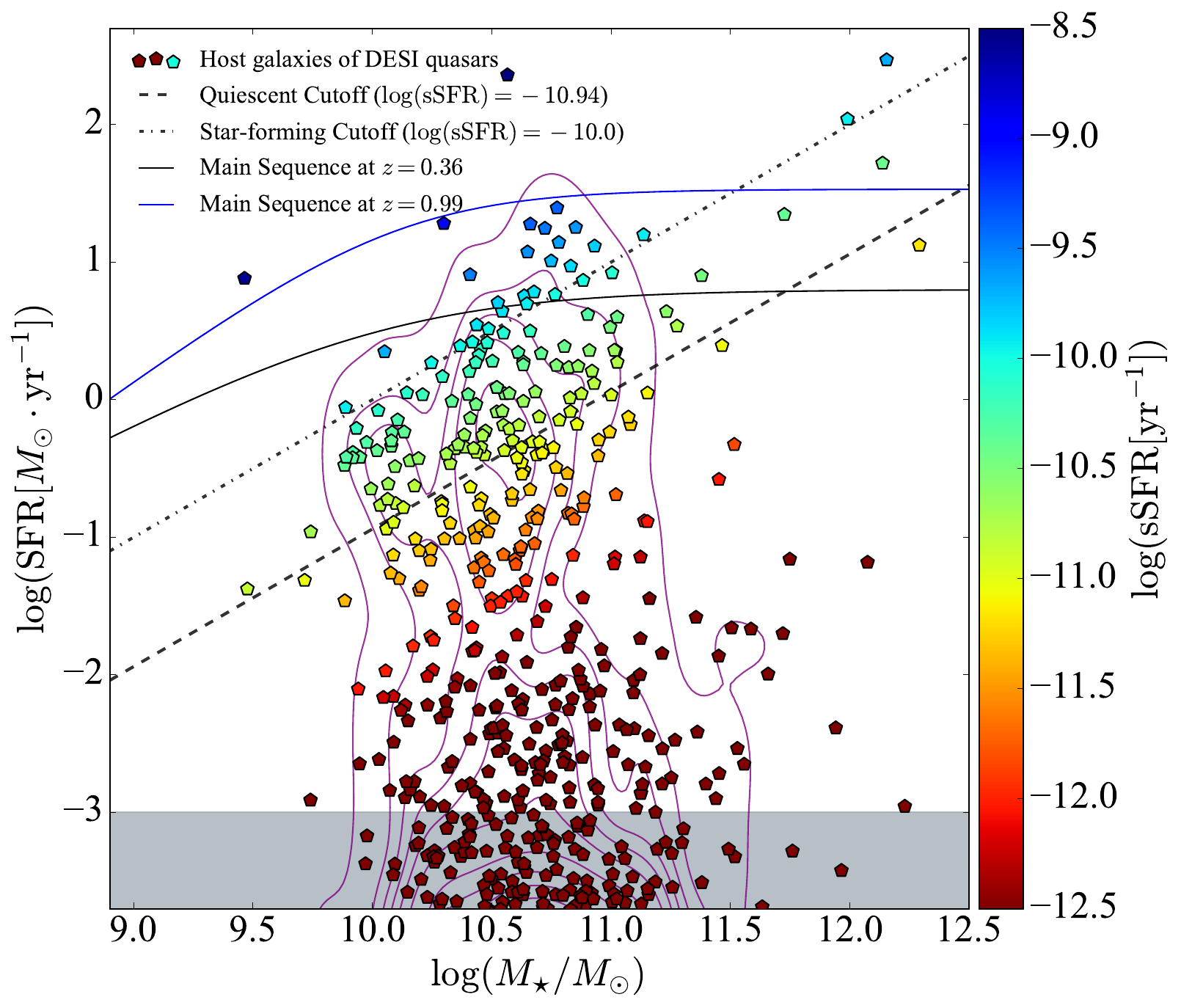}
    \caption{Distribution of our quasars in the $\mathrm{SFR} - M_{\star}$ plane. \myrev{
The overlaid black contours represent the 2D kernel density estimate, highlighting a distinct bimodal separation within the host galaxy population.
The data points in the shaded region only indicate upper limits.
We plot the star-forming main sequence for the galaxies at $z=0.36$ and $z=0.9$ from previous studies of \citet{SF_main_sequence_2023, SF_main_sequence_2019} for reference.
The majority of our host galaxies are quiescent with old stellar populations, clustering deeply into the red sequence ($\log(\mathrm{sSFR}) \lesssim -11.5$).
The secondary peak represents a transitioning green valley population ($-11.5 < \log(\mathrm{sSFR}) < -10.0$), while actively star-forming main sequence galaxies are notably absent.}
}
    \label{fig:main-sequence host gal}
\end{figure}

Using the SFR and $M_{\star}$ measured from the decomposed host galaxy spectra, we probe the star-forming properties of the host galaxies. The majority of the host galaxies are not actively forming stars, with low specific star formation rates ($\mathrm{sSFR} = \mathrm{SFR}/M_{\star}$). As seen in Figure \ref{fig:main-sequence host gal}, the sSFRs for our DESI quasars are typically below $-10.5$. Only a few of them are star-forming galaxies. The galaxy green valley is clearly observed between $-1.5<\log( \mathrm{SFR / M_{\odot} \cdot \mathrm{yr}^{-1}})<-0.5$ in Figure \ref{fig:main-sequence host gal}.  
\myrev{As highlighted by the density contours in Figure \ref{fig:main-sequence host gal}, our sample exhibits a distinct bimodal density structure. However, unlike standard galaxy populations, this bimodality does not bridge the main sequence and the red sequence. Instead, the upper density peak, spanning $-11.5 < \log(\mathrm{sSFR} [\mathrm{yr}^{-1}]) < -10.0$, corresponds directly to the transitioning green valley. This aligns with the prevalence of post-starburst host galaxies discussed in Section \ref{sec:post-starburst}. The lower density peak, separated by a localized density minimum at $\log(\mathrm{sSFR} [\mathrm{yr}^{-1}]) \approx -11.5$, consists of quiescent galaxies, many of which approach our SFR measurement limits (shaded region). The absence of galaxies on the star-forming main sequence \citep{SF_main_sequence_2019, SF_main_sequence_2023} provides further evidence that our extended quasar hosts represent an older, rapidly quenching, or already quenched population.}

\begin{figure}
    \centering
    \includegraphics[width=0.99\linewidth]{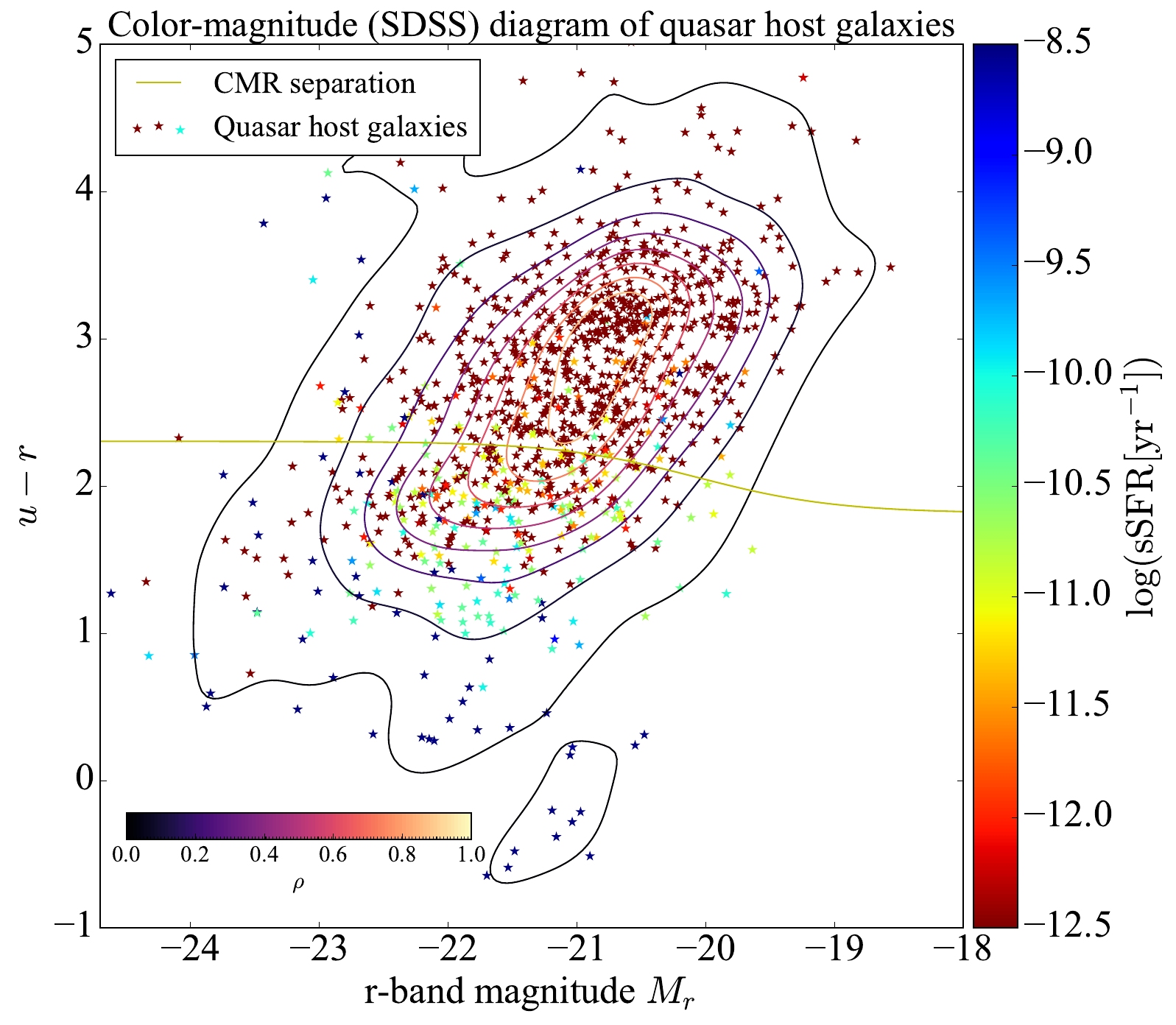}
    \caption{The $u-r$ vs. $r$ CMD for our quasar host galaxies. The individual galaxies are color-coded by their sSFRs. 
    The red line indicates the optimal color division for red galaxies and blue star-forming galaxies adopted from \citet{CMR_sep_2004ApJ...600..681B}.
    The red sequence and blue cloud are well-separated in the plane.  
    Only a small fraction of our host galaxies are star-forming.}
    \label{fig:CMD_u-r}
\end{figure}

In Figure \ref{fig:CMD_u-r}, we display the color magnitude diagram (CMD) for our quasar host galaxies. The magnitudes are calculated from the spectra using the SDSS filters, so that the results can be directly compared with previous results \citep{2004ApJ...600L..11B_Bell_2004a, 2004ApJ...608..752B_Bell_2004b, CMR_sep_2004ApJ...600..681B}.
We use the color division for red and blue galaxies from \citet{CMR_sep_2004ApJ...600..681B}, 
\begin{equation}
C_{ur} \left( M_r \right) = 2.06 - 0.244 \, \mathrm{tanh} \left[ \frac{M_r+20.07}{1.09} \right] .
\end{equation}
From their red continuum colors, we conclude that the majority of our decomposed host galaxy components have old stellar populations, which is consistent with their low star formation level as discussed above. 
Similar results have been found in quasar host galaxies with clear detection of stellar absorption lines \citep[e.g.,][]{YM_shenyue_2015ApJ...811...91M}.
Combining the stellar population from the color separation and sSFR measurements, we find that only a minority of our quasar hosts are star-forming.

\begin{figure}
    \centering
    \includegraphics[width=0.99\linewidth]{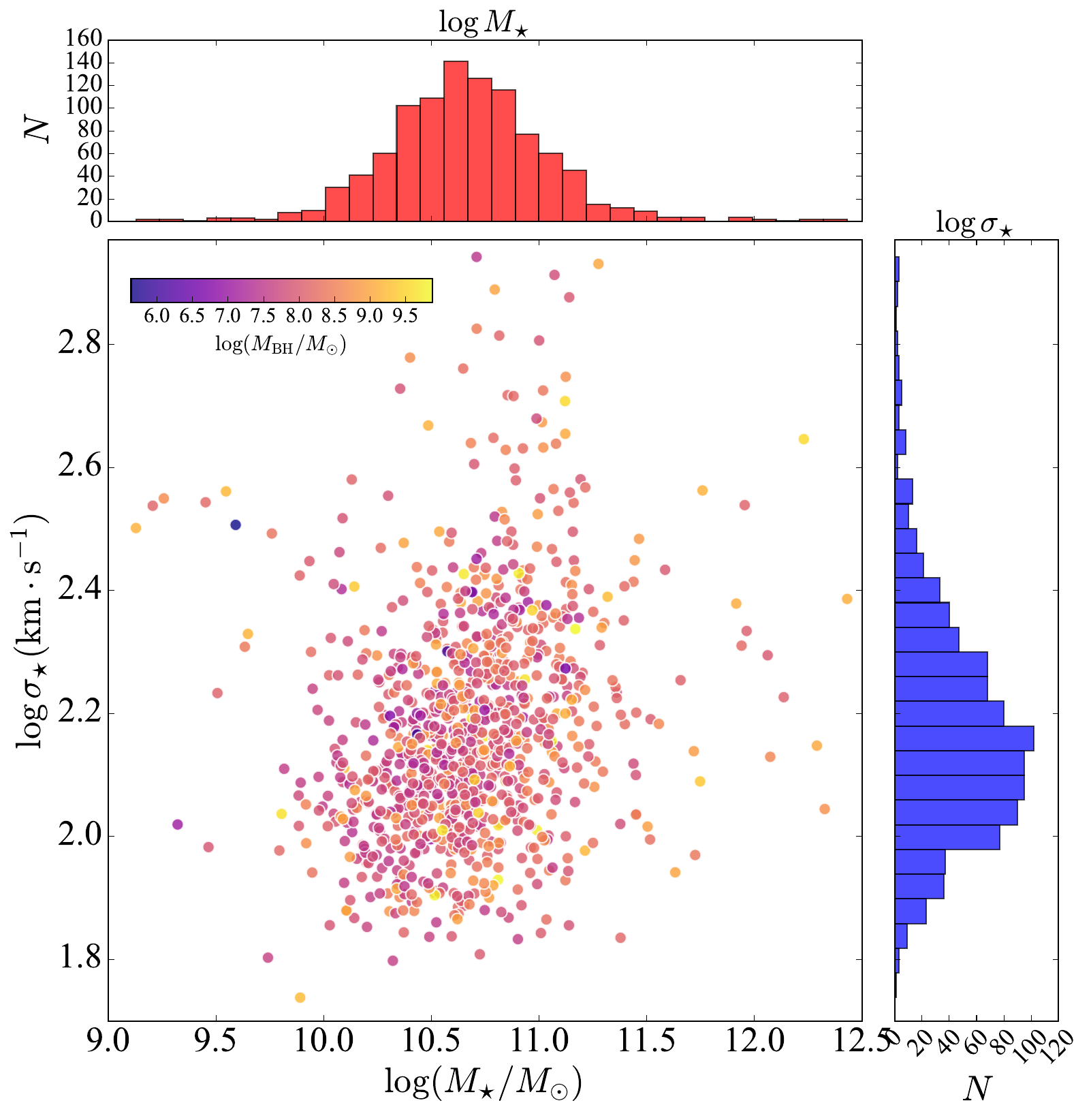}
    \caption{Distribution of our DESI quasars in the $M_{\star} - \sigma_{\star}$ plane, color-coded by the black hole mass. The majority of our DESI quasar hosts have relatively large stellar masses, and a wide spread range of $M_{\star}$ and  $\sigma_{\star}$.}
    \label{fig:mstar-sig}
\end{figure}

\myrev{Figure \ref{fig:mstar-sig} presents the distribution of our DESI quasars in the $M_{\star}$--$\sigma_{\star}$ plane. The stellar mass distribution shows a unimodal profile peaking at $M_{\star} \approx 10^{10.6} \, M_{\odot}$, while $\sigma_{\star}$ peaks at $\sim 150 \ \mathrm{km \ s^{-1}}$. For the general galaxy population, the volume-averaged number density is dominated by star-forming galaxies at the low-mass end. In contrast, the shape and the $\sim 10^{10.6} \, M_{\odot}$ peak of our sample closely match the stellar mass function of quiescent galaxies \citep[e.g.,][]{2013A&A...556A..55I, 2013ApJ...777...18M}. This massive host galaxy preference is consistent with previous quasar demographic studies \citep[e.g.,][]{2014ARA&A..52..589H, 2022ApJ...934..130Z}, and aligns with the post-starburst evolutionary scenario discussed in Section \ref{sec:post-starburst}.}

\subsection{Prevalence of post-starburst features}\label{sec:post-starburst}

Post-starburst galaxies are transitionary species of galaxies (time duration $\sim 300$ Myr) that are often characterized by their strong Balmer absorption lines and weak emission lines. They comprise less than 1\% of the local galaxy population and up to $\sim 5 \%$ of galaxies at $z \sim 2$
\citep[e.g.,][]{2011ApJ...741...77S, PSB_evolution_2016MNRAS.463..832W, PSB_manga_2019MNRAS.489.5709C}.
We adopt the selection criteria of post-starburst galaxies from \citet{2021ApJ...907L..21D} that require $\mathrm{H} \delta_{A} > 4$ \AA\ and no recent star formation ($\log \mathrm{sSFR} < -10.94$) for our quasar hosts. 
\myrev{To accurately apply this criterion and mitigate the severe degeneracy between AGN emission and host stellar absorption, we measure the $\mathrm{H}\delta_A$ equivalent width on the decomposed host galaxy spectra. The calculation follows the standard Lick $\mathrm{H}\delta_A$ index definition \citep{1997ApJS..111..377W}. The local pseudo-continuum is determined via a linear interpolation between the mean fluxes of the blue ($4041.60 - 4079.75$ \AA) and red ($4128.50 - 4161.00$ \AA) sidebands. The absorption equivalent width is then integrated over the central feature bandpass ($4083.50 - 4122.25$ \AA). 
This empirical approach, applied post-decomposition, effectively circumvents the AGN degeneracy without heavily relying on specific stellar population template priors}

\begin{figure}
    \centering
    \begin{minipage}{0.5\textwidth}
        \centering
        \includegraphics[width=1\textwidth]{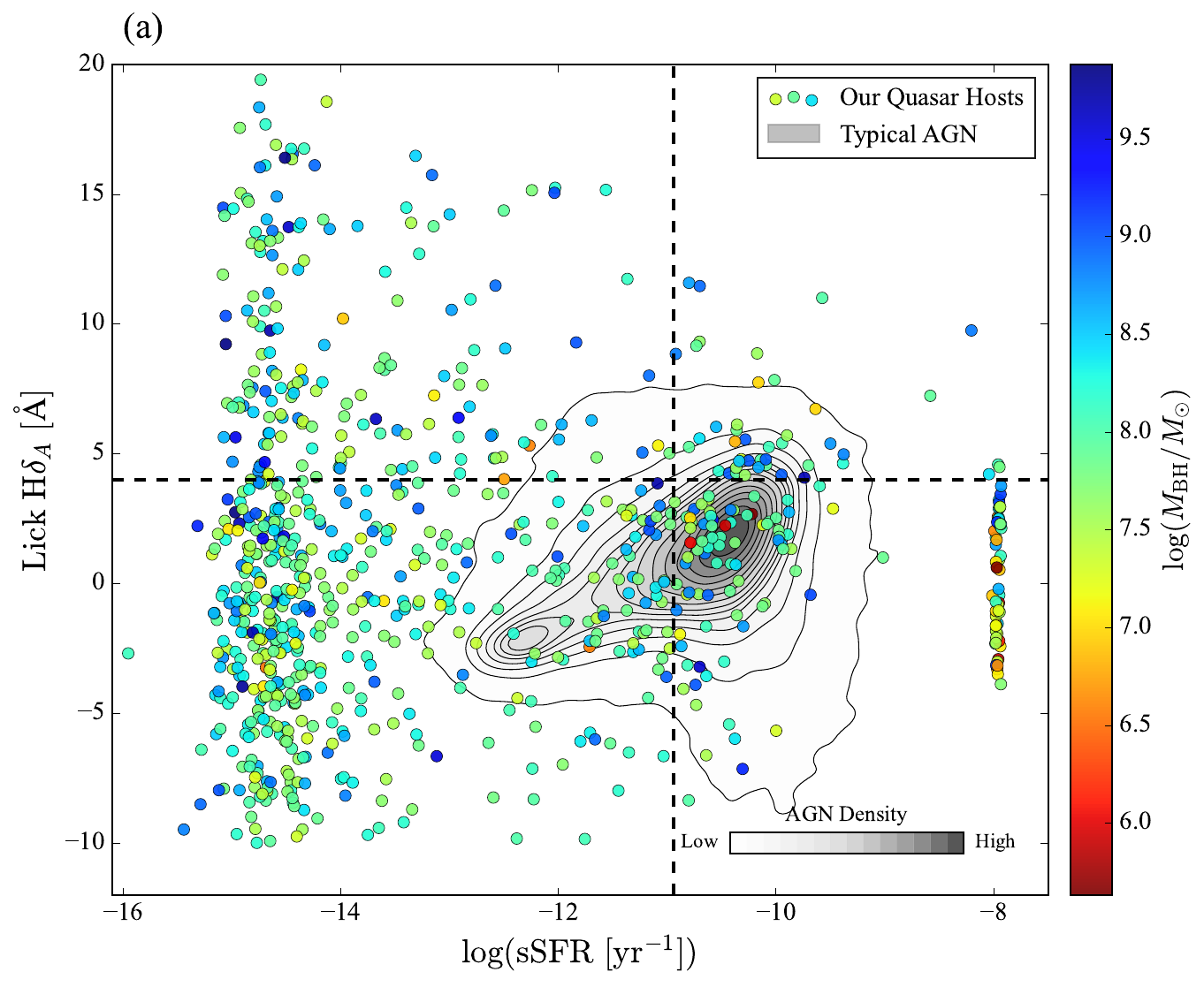}
    \end{minipage}%
    \hfill %
    \begin{minipage}{0.5\textwidth}
        \centering
        \includegraphics[width=0.9\textwidth]{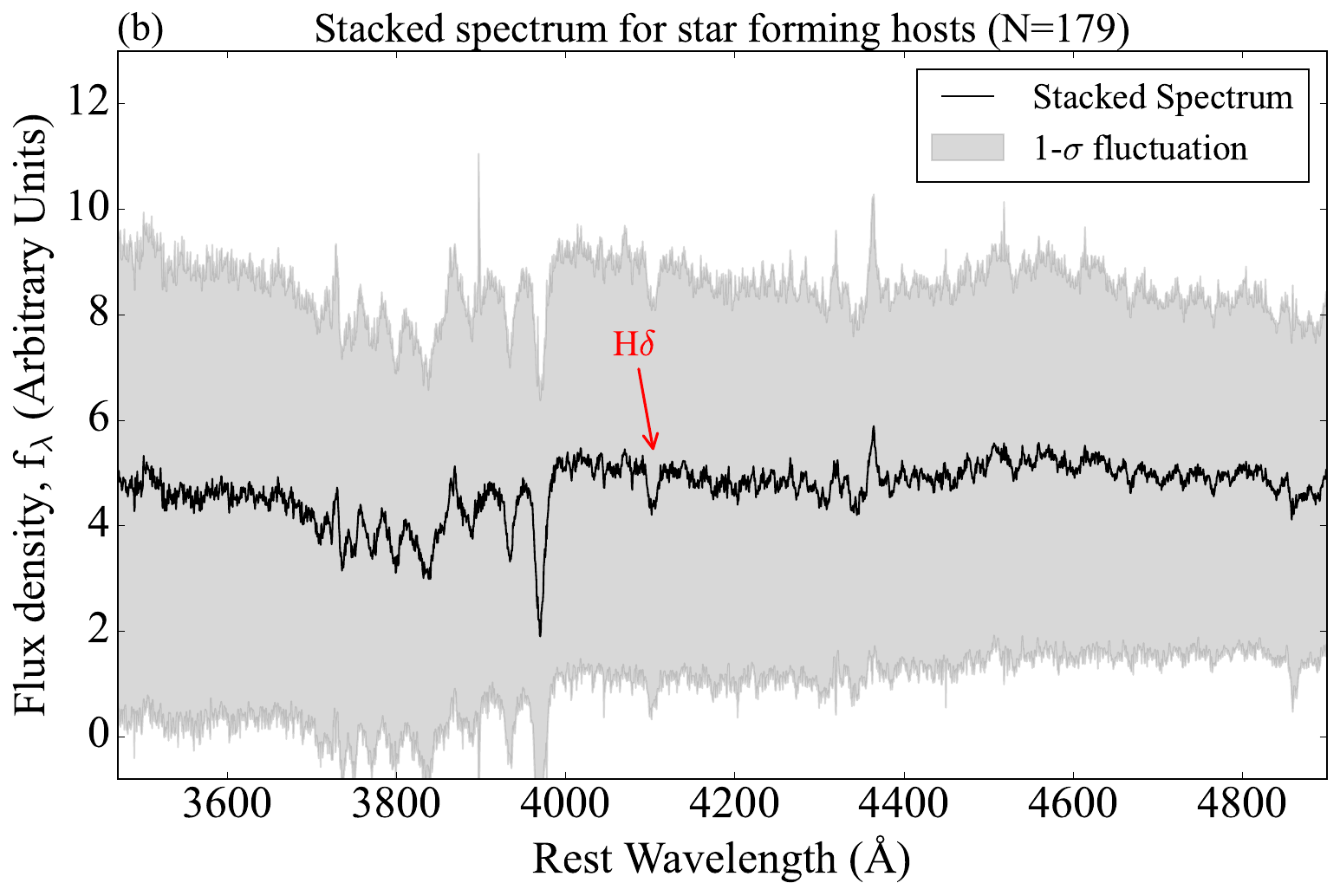}
    \end{minipage}
    \hfill%
     \begin{minipage}{0.5\textwidth}
        \centering
        \includegraphics[width=0.9\textwidth]{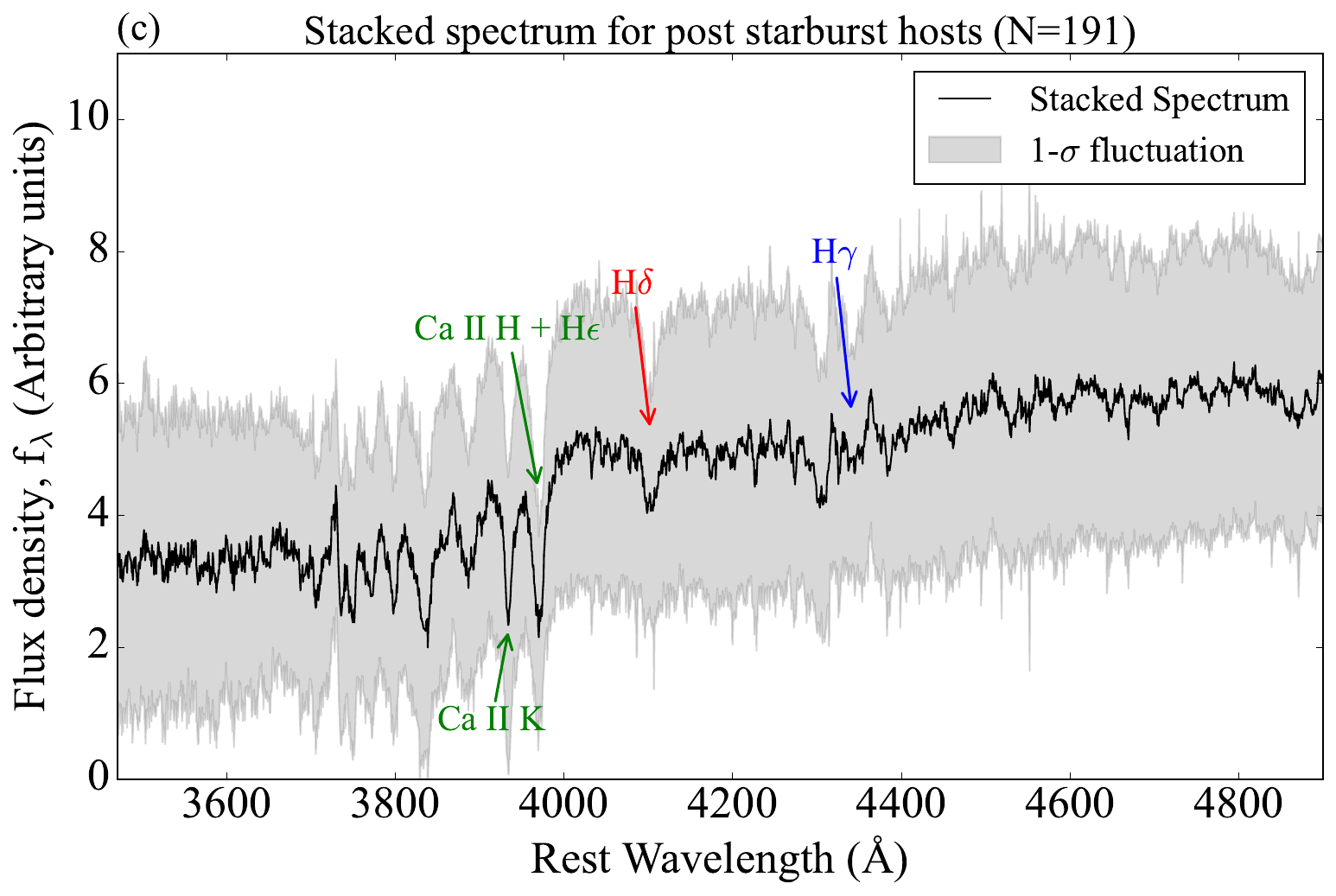}
    \end{minipage}
    \vspace{0.5em} %
    \caption{\myrev{Absorption line features and sSFRs. \textit{Upper panel:} Distribution of our quasar hosts in the $\log(\mathrm{sSFR})$ versus $\mathrm{H}\delta_A$ plane. Data points are color-coded by black hole mass ($\log M_{\mathrm{BH}}/M_{\odot}$). Grey contours show the density distribution of an SDSS AGN control sample (MPA-JHU catalog). The horizontal and vertical dashed lines ($\mathrm{H} \delta_A = 4$ \text{\AA} and $\log \mathrm{sSFR} = -10.94$) define the post-starburst selection region. About 23\% of our quasar hosts are post-starburst galaxies. \textit{Middle} and \textit{lower panels:} Stacked spectra of the star-forming and post-starburst host galaxies, respectively.}}
    \label{fig:post_starburst}
\end{figure}

We present the statistical analysis of post-starburst quasar host galaxies in Figure \ref{fig:post_starburst}. 
The two separation criteria are displayed by the dashed black lines in the upper panel. 
In the middle and lower panels of Figure \ref{fig:post_starburst}, we present the composite spectra of star-forming and post-starburst galaxies constructed by the median stacking technique. The standard deviations for the spectra are shown by the grey shaded regions.

\myrev{The prevalence of post-starburst features is observed in our decomposed host galaxies. As shown in Figure \ref{fig:post_starburst}(a), approximately 23\% of our extended quasar hosts fall into the post-starburst regime. To properly contextualize this result, we compare our sample against a baseline of typical AGN host galaxies.}

\myrev{We constructed a control sample of typical AGNs using the Sloan Digital Sky Survey (SDSS) Data Release 16 \citep[DR16;][]{2020ApJS..249....3A} and MPA-JHU value-added catalog \citep{2003MNRAS.341...33K, 2004MNRAS.351.1151B}. We selected sources confidently classified as AGNs based on the BPT diagram \citep[e.g., Seyferts and LINERs;][]{2004MNRAS.351.1151B} at $z < 0.3$. The density distribution of this control sample is displayed as grey background contours in Figure \ref{fig:post_starburst}(a), sharing the same parameter space ($\log(\mathrm{sSFR})$ vs. $\mathrm{H}\delta_A$) typically utilized in recent literature \citep[e.g.,][]{2021ApJ...907L..21D}.}

\myrev{As the contours indicate, typical optically-selected AGNs predominantly occupy the star-forming main sequence or the green valley, generally exhibiting weak $\mathrm{H} \delta_{A}$ absorption. Applying our strict selection criteria ($\mathrm{H} \delta_{A} > 4$ \AA\ and $\log \mathrm{sSFR} < -10.94$) to this control sample yields a post-starburst fraction of only $\sim 1.2\%$. This minimal fraction is consistent with previous demographic studies, which establish that post-starburst galaxies constitute $\lesssim 1\%$ of the general local galaxy population \citep[e.g.,][]{1996ApJ...466..104Z, 2007MNRAS.381..187G} and typically only a few percent of normal AGN and quasar hosts \citep[e.g.,][]{2013ApJ...762...90C, 2014ApJ...792...84Y}. Against this established literature consensus and our direct AGN baseline, the 23\% fraction observed in our study is notably elevated. This robust comparison supports the scenario that our morphological criteria—specifically the requirement for extended optical morphologies—effectively pre-selects galaxies with large effective radii, characteristics frequently associated with this transitional post-starburst phase.}

In Figure \ref{fig:post_starburst}(b), the composite spectrum of post-starburst hosts demonstrates strong Balmer absorption series and \caii\ HK lines. As an example of individual post-starburst quasar hosts, we refer to the two cases in Figure \ref{fig:decomposed_host_galaxies_in_detail}. Balmer absorption series, together with \caii\ HK and \mgi\ absorption lines, are strong evidences of their post-starburst activities.
The high fraction of post-starburst galaxies in our sample likely result from the selection criterion that requires extended morphology, since the post-starburst galaxies typically have extended shapes and large effective radius $R_{\mathrm{e}}$. As a possible explanation of the post-starburst host scenario, it is natural to observe AGNs living inside post-starburst galaxy environments if their precursors reside in actively star-forming galaxies at higher redshifts, as has been revealed in the case at $z \sim 7$ \citep[e.g.,][]{2021ApJ...907L...1W}. After the rapid concurrent growth of SMBHs and their massive host galaxies, both would follow the evolutionary track of AGN quenching and end up with a stage of post-starburst galaxies hosting SMBHs. Thus, 23\% of our DESI quasars went through rapid growth of SMBHs and their host galaxies, and eventually reach the observed post-starburst host galaxies.

\subsection{Unique properties of our extended quasars}

In the earlier analyses of our quasars, we have probed a large parameter space in black hole mass, host stellar mass, and host galaxy SFR. The black hole mass $M_{\mathrm{BH}}$ ranges from $10^{5.6 - 9.9} M_{\odot}$, stellar mass $M_{\star}$ ranges from $10^{9.1 - 12.4} M_{\odot}$, SFR ranges from $10^{-4}$ to $10^{3.5} \, M_{\odot}\cdot \rm yr^{-1}$ in our sample. The median value of $M_{\mathrm{BH}}$ is $8.02 \, M_{\odot}$, the median $M_{\star}$ value is $10.64 \, M_{\odot}$, and the median SFR value is $10^{-1.9} \, M_{\odot}\cdot \rm yr^{-1}$.

Our results reveal a previously less-studied population of active SMBHs hosted by quiescent galaxies. We consider the host galaxies with $\log \left( \mathrm{sSFR} \left[\mathrm{yr}^{-1} \right] \right) < -10.94$ as quiescent galaxies. The fraction of quiescent host galaxies in our sample is around 83\%. Their host stellar populations are old, with inactive star formation (mainly below $\mathrm{SFR} < 10 \, M_{\odot} \cdot \mathrm{yr}^{-1}$, see Section \ref{sec:host_gal_prop}). 
\myrev{This is very different from previous findings of vigorous star-formation in low-$z$ quasar hosts \citep[e.g.,][]{2021ApJ...910..124X, 2022ApJ...934..130Z}. This apparent discrepancy arises from a strong selection effect introduced by our morphological requirement. While previous studies typically included or focused on compact quasars, our requirement for extended optical morphologies systematically biases our sample toward quasars hosted by massive galaxies with large effective radii and prominent spheroidal components. It is a well-established empirical relation in galaxy evolution that such structurally evolved, bulge-dominated galaxies are strongly correlated with lower star formation activity and a higher likelihood of being quenched \citep[e.g.,][]{2003MNRAS.341...54K, 2012ApJ...753..167B, 2013ApJ...776...63F}. Consequently, our morphological selection naturally pre-selects a population of hosts that inherently exhibit lower SFRs.}

Due to the flux-limited nature of our quasar sample, the selected quasars are all relatively bright, presumably with high accretion rates. Thus, as a complementary result to previous studies \citep{2021ApJ...910..124X, 2022ApJ...934..130Z, 2023ApJ...944...30M}, our sample shows a negative feedback that SMBHs exerted on their host galaxies.
Combining the post-starburst quasar host galaxy distribution, the SFH results, and the inactive recent star formation behaviors, we find that AGNs started to quench star formation in these quasars at $z>2$ (up to $z \sim 4-6$). The early AGN quenching leads to the observed $0.1<z<1$ quiescent host galaxies, and some of them are post-starburst galaxies.

\begin{figure}
    \centering
    \begin{minipage}{0.5\textwidth}
        \centering
        \includegraphics[width=\textwidth]{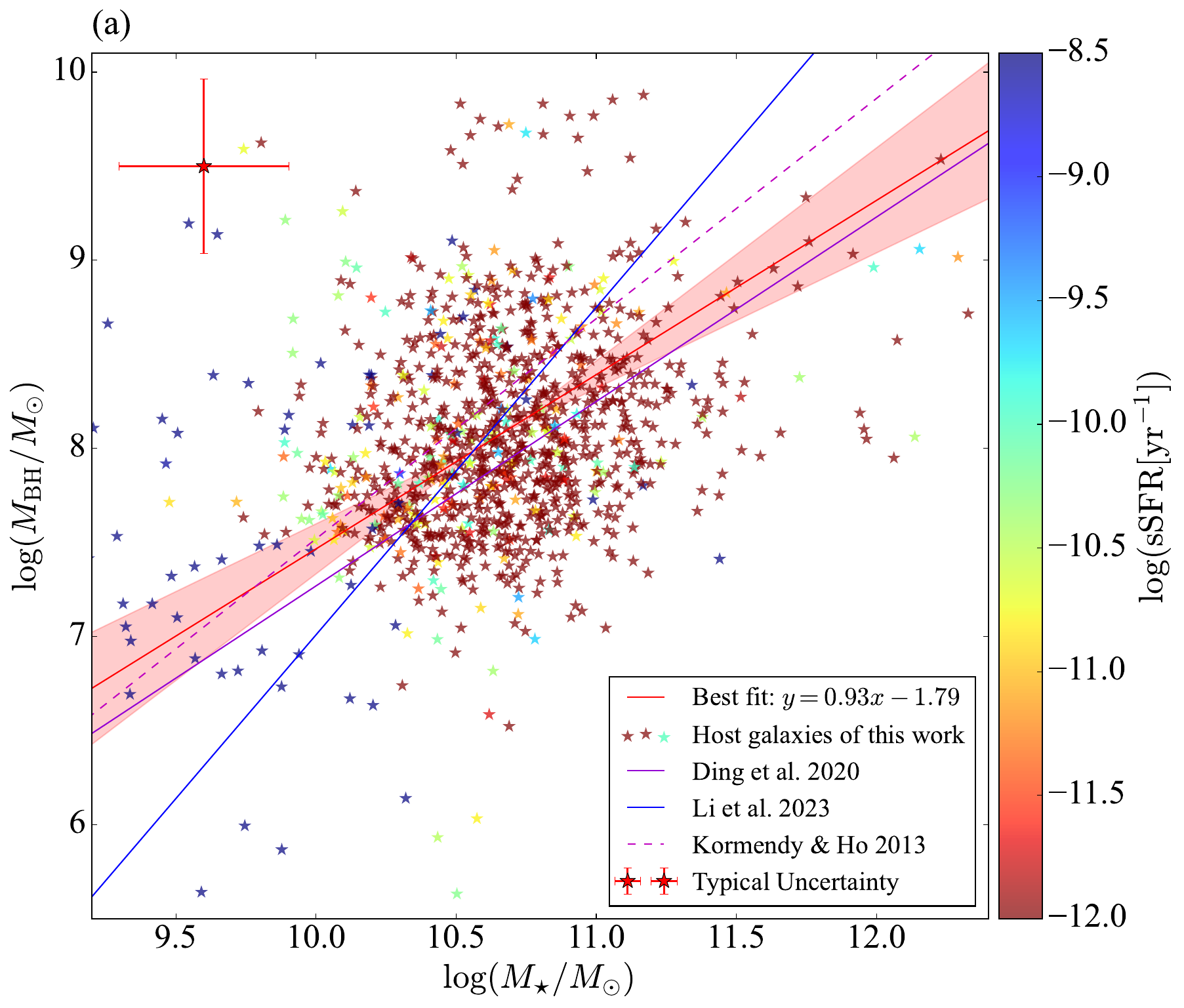}
    \end{minipage}%
    \hfill %
    \begin{minipage}{0.5\textwidth}
        \centering
        \includegraphics[width=\textwidth]{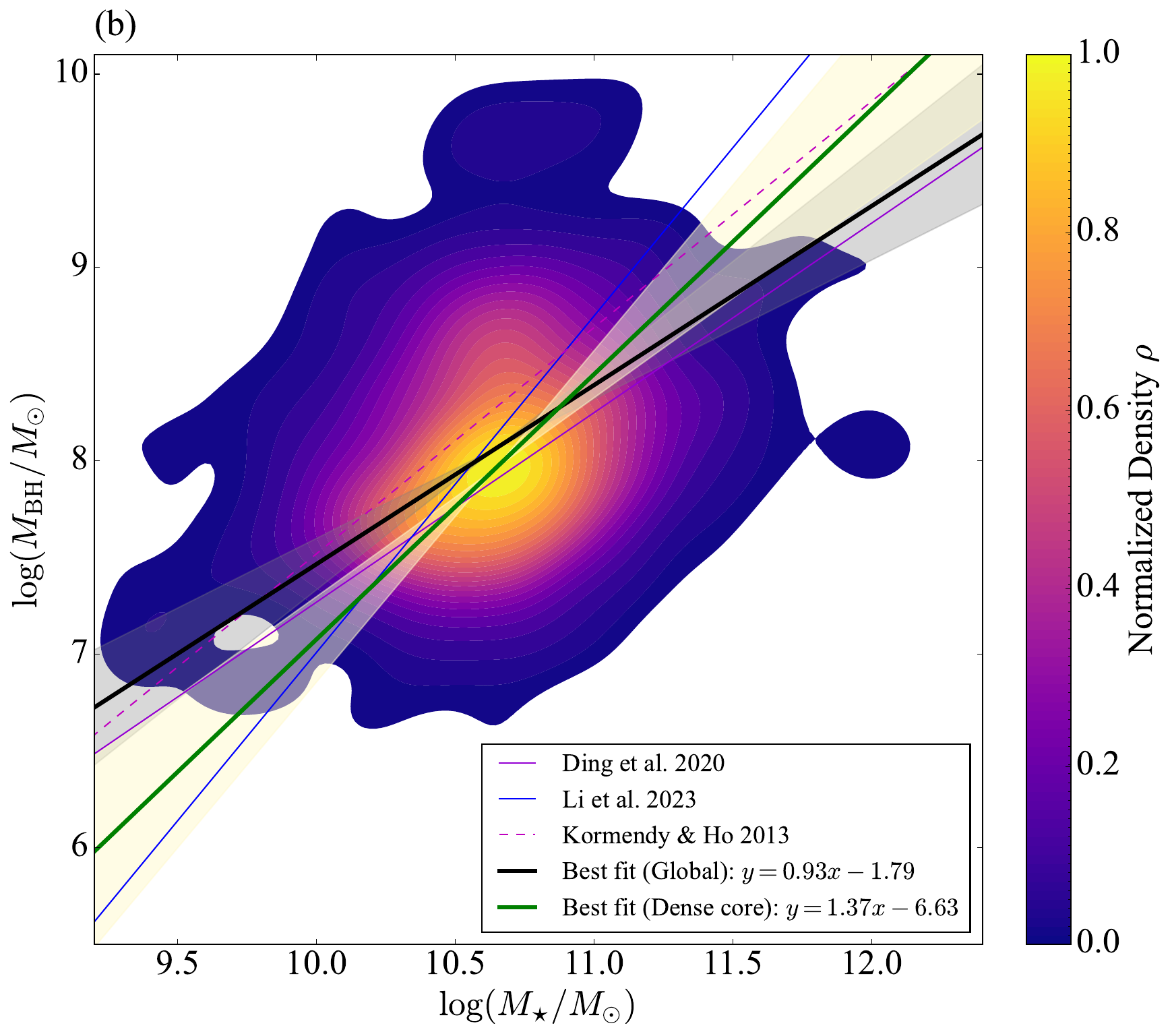}
    \end{minipage}
    \vspace{0.5em} %
    \caption{The $M_{\mathrm{BH}} - M_{{\star}}$ relation for our sample of $1083$ quasars with extended morphology. The upper panel is color-coded by sSFR, and the lower panel is color-coded by the density of the quasar distribution. The solid purple line shows the relation at $1<z<2$ from \citet{2020ApJ...888...37D_Mbh_Mstar_Ding2020}, the blue line represents the relation at $0.2<z<0.8$ from \citet{2023ApJ...954..173L_Li_and_shenyue} for 38 broad-line quasars, and the dashed magenta line shows the local relation from \citet{2013ARA&A..51..511K}. 
      \myrev{In the upper panel, red error bar in the upper left indicate the typical $1\,\sigma$ uncertainties, which include systematic errors. The red solid line and shaded region show the global ODR fit. In the lower panel, the green line and shaded region represent the ODR fit restricted to the high-density core (617 quasars). It has a steeper slope compared to the whole population that is displayed by the black solid line.}}
    \label{fig:Mbh_Mstellar}
\end{figure}

\section{Discussion}\label{sec:discussion}

We examine the correlations between the physical properties of black holes and their host galaxies within the redshift range $0.1<z<1$, and compare with the well-established local scaling relations by \citet{2013ARA&A..51..511K}. We further discuss evidence of co-evolution between quasars and their host galaxies. We exclude about 4\% of our quasars that were not well modeled. The final sample size is 1083. Due to our sample selection, the stellar masses are mostly around $\log(M_{\star}/M_{\odot}) \approx 10.0 \, \text{--} \, 11.3 $, and the black hole masses are mostly around $\log(M_{\mathrm{BH}}/M_{\odot}) \approx 7.1 \, \text{--} \, 9.1 $. 

\subsection{$M_{\mathrm{BH}} - M_{\star}$ relation} \label{subsec:M-bh_stellar relation}

In Figure \ref{fig:Mbh_Mstellar}, we show the distribution of our quasars in the $M_{\mathrm{BH}} - M_{\star}$ plane. To account for the measurement uncertainties in both variables, we employ the Orthogonal Distance Regression (ODR) method. Furthermore, to avoid an artificially shallow slope caused by regression dilution, we incorporate physically motivated systematic uncertainties ($\sigma_{\mathrm{sys}, M_{\star}} = 0.2$ dex and $\sigma_{\mathrm{sys}, M_{\mathrm{BH}}} = 0.3$ dex) in quadrature with the formal statistical errors. 
The resulting global best-fit linear relation is $\log(M_{\mathrm{BH}}/M_{\odot}) = 0.93 \times \log(M_{\star}/M_{\odot}) - 1.79$. 
This slope is broadly consistent with the canonical local relations \citep[typically $\sim 1.1-1.2$; e.g.,][]{2013ARA&A..51..511K} within the $1\sigma$ uncertainties.
As shown in Figure \ref{fig:Mbh_Mstellar}, our quasars do not exhibit a severe bias towards over-massive black holes at the low-mass end; rather, they are evenly scattered around the local empirical relations. The minor difference between our fitted slope and the canonical value is primarily a consequence of our sample selection. Because we select bright quasars with extended optical morphologies, our sample inherently favors massive, often quiescent, host galaxies, which restricts the dynamic range of the stellar mass. Fitting a linear relation over such a restricted dynamic range in the presence of large intrinsic scatter naturally dilutes the recovered slope \citep{2007ApJ...670..249L, 2010ApJ...713...41S}.

\myrev{
Despite a large scatter around the global fitted correlation, Figure \ref{fig:Mbh_Mstellar} (a) shows that quasars with quiescent and star-forming host galaxies are evenly distributed around the fit, suggesting that the sSFR of host galaxies does not largely dictate the distribution of quasars in this plane. To mitigate the impact of extreme outliers and selection biases at the sample margins, we utilize a subsample strictly restricted to the high-density regime in the $M_{\mathrm{BH}} - M_{\star}$ plane (the dense core). For this dense core, the best-fit ODR relation dramatically steepens to $\log(M_{\mathrm{BH}}/M_{\odot}) = 1.37 \times \log(M_{\star}/M_{\odot}) - 6.63$ (see Figure \ref{fig:Mbh_Mstellar} (b)). This core slope is highly consistent with the canonical steep empirical relations, demonstrating that the underlying intrinsic relation is indeed consistent with local massive galaxies.
}

\begin{figure}
    \centering
    \begin{minipage}{0.45\textwidth}
        \centering
        \includegraphics[width=\textwidth]{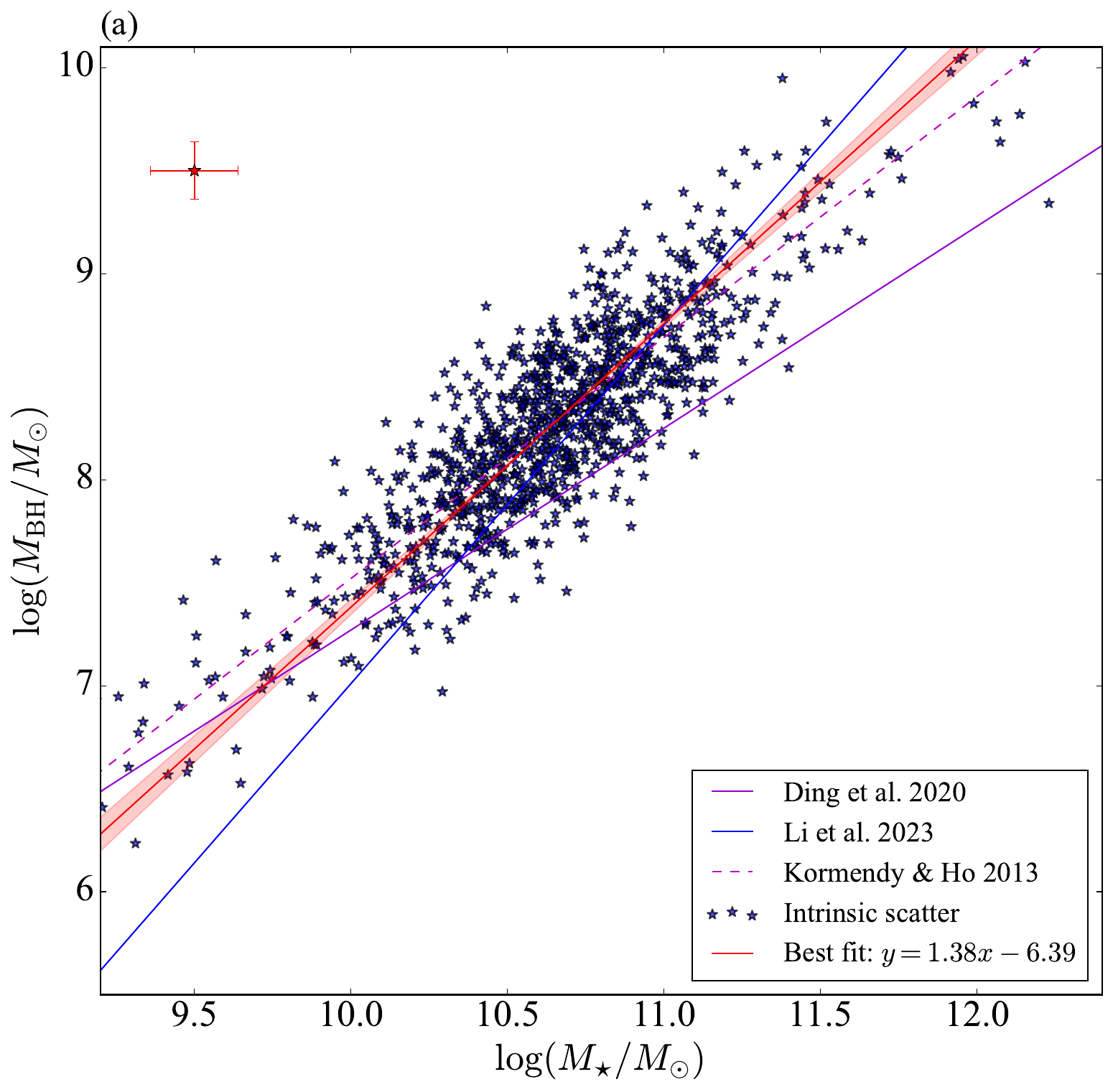}
    \end{minipage}%
    \hfill %
    \begin{minipage}{0.45\textwidth}
        \centering
        \includegraphics[width=\textwidth]{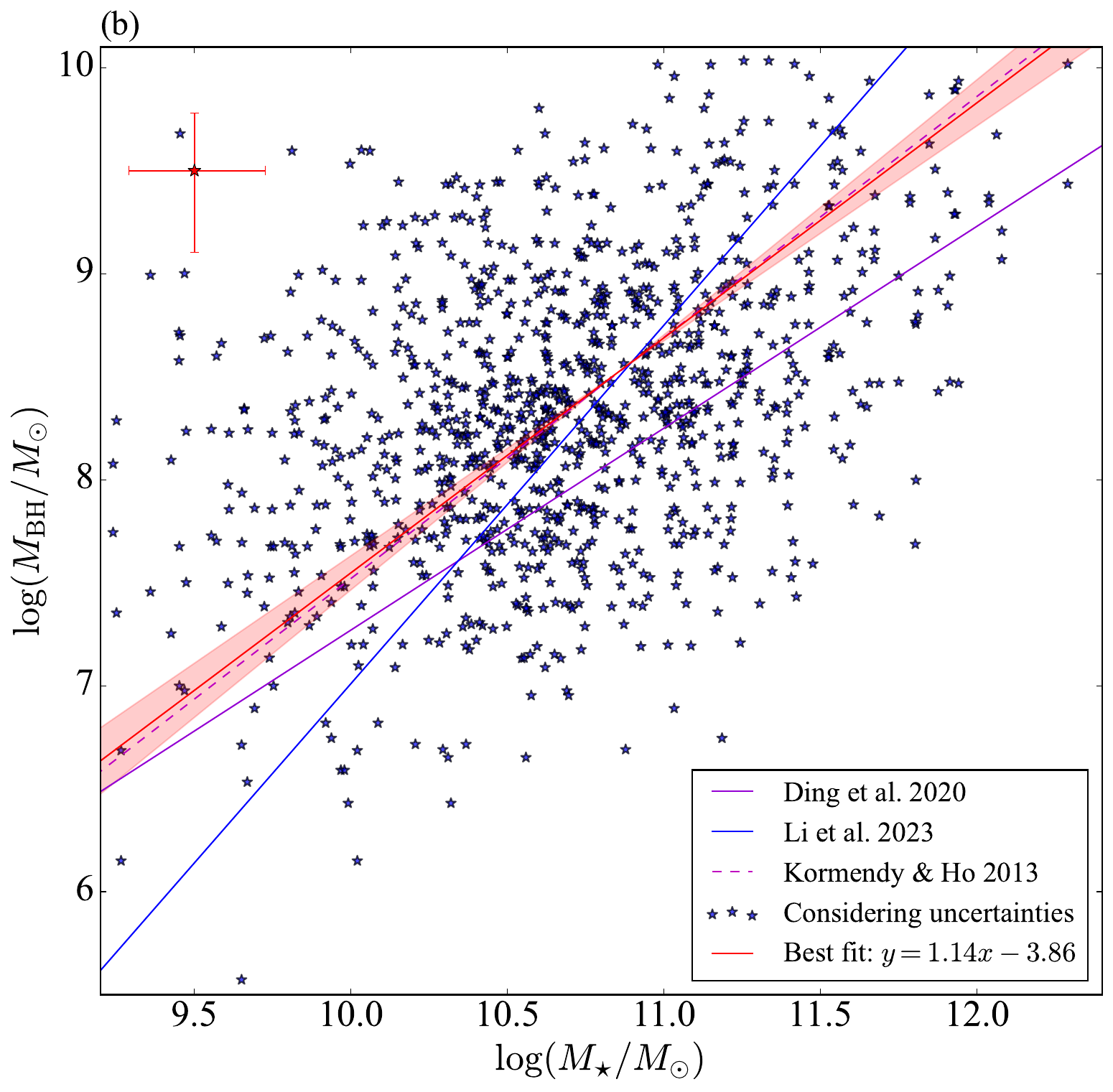}
    \end{minipage}
    \vspace{0.5em} %
    \caption{$M_{\mathrm{BH}} - M_{\star}$ distribution for a sample of mock galaxies that follow the local relation. The upper panel only considers the intrinsic scatter/uncertainty, and the lower panel includes all the measurement uncertainties. Notations are the same as those in Figure \ref{fig:Mbh_Mstellar}. Despite the large scatter in the lower panel, the best-fit relation is consistent with that in the upper panel, suggesting that the large scatter in Figure \ref{fig:Mbh_Mstellar} does not affect our result.
    }
    \label{fig:Mbh_Mstellar_err}
\end{figure}

\myrev{
We further investigate the influence of different uncertainties on the observed scaling relation, displaying the mock test results in Figure \ref{fig:Mbh_Mstellar_err}. In the upper panel, we generate a sample of mock galaxies following the local relation, utilizing the same stellar mass distribution as our quasar host galaxies and incorporating a Gaussian intrinsic scatter of 0.28 dex \citep{2013ARA&A..51..511K}. In the lower panel, we inject the uncertainties introduced by our pipeline (photometry, joint decomposition, and physical property measurements), resulting in a total uncertainty level of $\sim 0.7$ dex. After injecting these large uncertainties, the scaling relation becomes heavily obscured and visually much less obvious. 
}

\myrev{
This mock test provides crucial empirical context for our actual observations in Figure \ref{fig:Mbh_Mstellar}: it demonstrates that the presence of large, under-characterized uncertainties in high-redshift samples can severely flatten and mask the intrinsic steep relation. Our empirical ODR fitting directly corroborates this mock test—when statistical errors were used alone, the relation appeared artificially shallow, but once systematic uncertainties were properly modeled, the steep intrinsic relations ($0.93$ for global, $1.37$ for the dense core) were successfully recovered. Therefore, contrary to a superficially shallow observed slope, our quasars with extended morphologies inherently follow a steep $M_{\mathrm{BH}} - M_{\star}$ relation consistent with local demographics, but this relation is heavily modulated by observational uncertainties and flux-limited selection effects. }
 
\subsection{$M_{\mathrm{BH}} - \sigma_{\star}$ relation} \label{subsec:M-sigma relation}

\begin{figure}
    \centering
    \begin{minipage}{0.5\textwidth}
        \centering
        \label{Mbh_sigma_whole}
        \includegraphics[width=\textwidth]{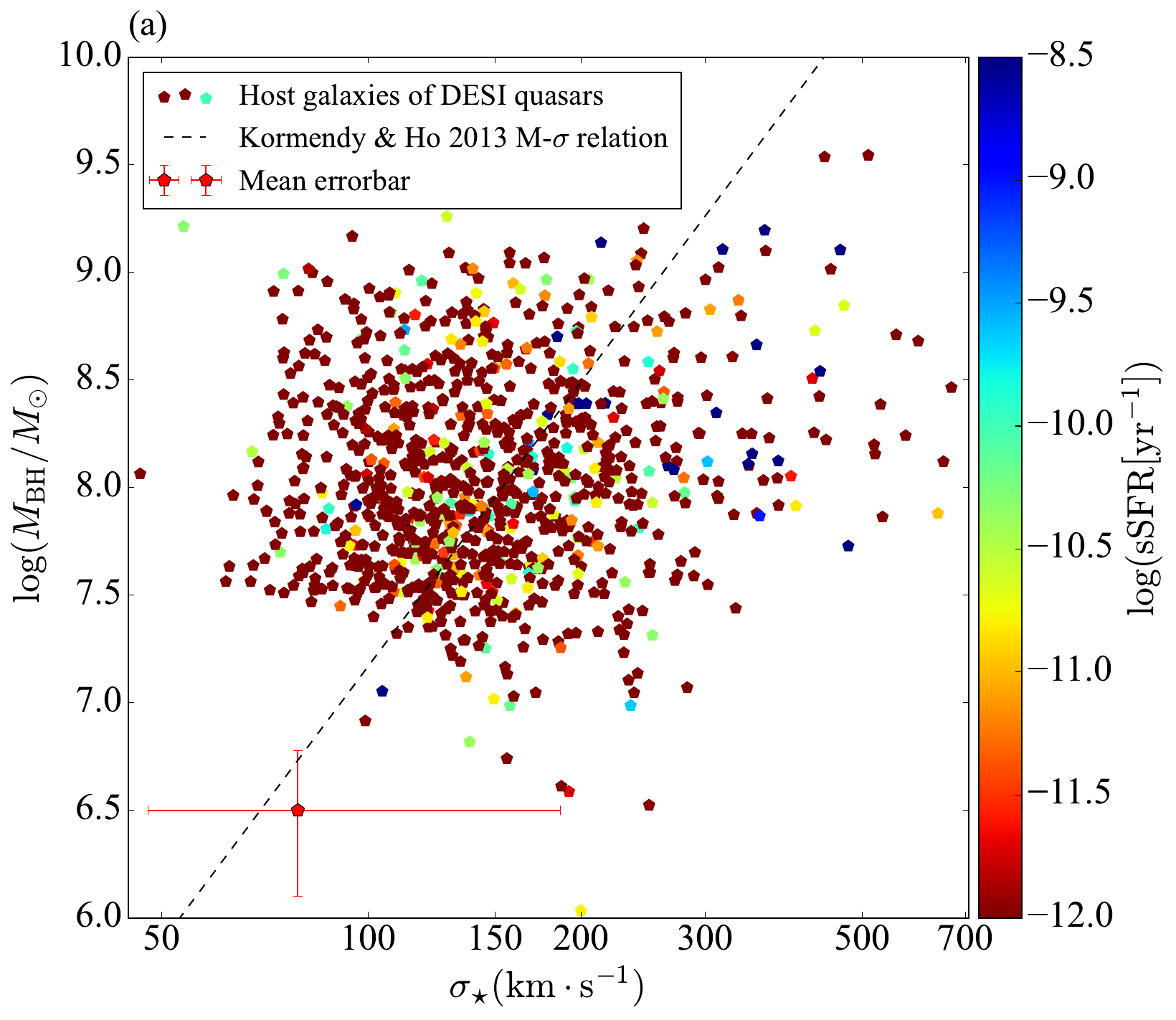}
    \end{minipage}%
    \hfill %
    \begin{minipage}{0.5\textwidth}
        \centering
        \label{Mbh_sigma_better_sigma}
        \includegraphics[width=\textwidth]{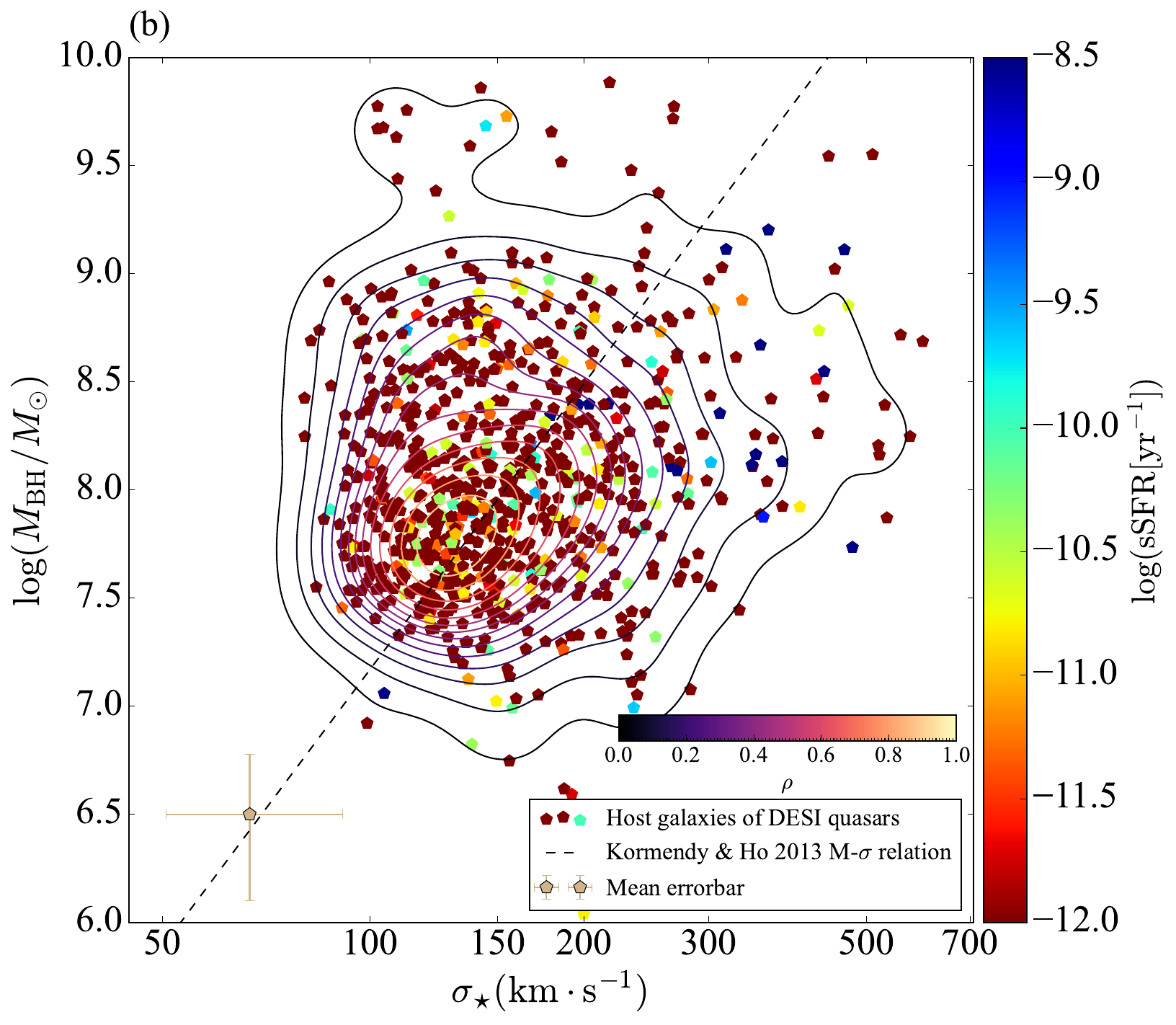}
    \end{minipage}
    \vspace{0.5em} %
    \caption{$M_{\mathrm{BH}} - \sigma_{\star}$ relation for our quasars. The pentacles show our quasar sample color-coded by sSFR of the host galaxies. The mean errors are plotted in the bottom left corner of the two panels.    
    We plot the classical $M_{\mathrm{BH}} - \sigma_{\star}$ relation for local dormant galaxies from \citet{2013ARA&A..51..511K} as reference.  
    In the upper panel, the whole sample reveals no significant correlation between $M_{\mathrm{BH}}$ and $\sigma_{\star}$.
    The lower panel shows a subsample with relatively more reliable $\sigma_{\star}$ estimations (errors smaller than 0.6 dex). 
    The density contours of this distribution is plotted. This subsample still show a large scatter around the local relation, and exhibit no significant correlation between $M_{\mathrm{BH}}$ and $\sigma_{\star}$.
    }
    \label{fig:M-sigma}
\end{figure}

We portray the distribution of our quasars in the $M_{\mathrm{BH}}$ - $\sigma_{\star}$ plane in Figure \ref{fig:M-sigma}. In the upper panel, we show the whole quasar sample. These quasars exhibit a large scatter around the local relation \citep{2013ARA&A..51..511K}. A linear regression fit to the data gives a Pearson correlation coefficient of $r = -0.14$, and p-value of $p=4.7 \times 10^{-6}$, suggesting that there is no significant linear correlation between $M_{\mathrm{BH}}$ and $\sigma_{\star}$. We also perform a linear fit to a subsample with better $\sigma_{\star}$ measurements (errors in $\sigma_{\star}$ smaller than 0.6 dex, the lower panel), and still find no significant linear correlation between $M_{\mathrm{BH}}$ and $\sigma_{\star}$. 
\myrev{We note that, similar to the $M_{\mathrm{BH}} - M_{\star}$ relation, the observed $M_{\mathrm{BH}}-\sigma_{\star}$ relation is severely affected by observational scatter. As demonstrated by our mock tests (See Figure \ref{fig:mock_vd_comparison}), the large measurement uncertainties associated with $\sigma_{\star}$ prevent us from deriving a robust intrinsic slope. Therefore, we present the $M_{\mathrm{BH}}-\sigma_{\star}$ fitting results primarily for completeness and reference, and refrain from drawing strong physical conclusions regarding its evolutionary trend.}

There are a few possible reasons for the non-correlation above, including the large measurement uncertainties of $M_{\mathrm{BH}}$ and $\sigma_{\star}$ and possible large scatters in the intrinsic relation between $M_{\mathrm{BH}}$ and $\sigma_{\star}$ \citep[e.g.][]{2015ApJ...805...96S_Shen2015_m_sigma}. The most likely reason is the large measurement uncertainties, as seen from the typical error bars in Figure \ref{fig:M-sigma}. Our mock test shows that the large uncertainties could have prevented us from obtaining an observed relation even if there is an intrinsic relation (similar to Figure \ref{fig:Mbh_Mstellar_err}). 
Another probable reason is that the velocity dispersion measured by \texttt{Bagpipes} is not the typical velocity dispersion around the central bulge, but rather an integration of all velocity dispersions inside a DESI fiber aperture, including those at the outer region of the host galaxy. In addition, some host galaxies probably do not have well-established virial bulges.
Therefore, based on the current data, we are not able to build a reliable $M_{\mathrm{BH}}$ and $\sigma_{\star}$ relation for our quasar host galaxies.

\section{Summary}\label{sec:summary}

In this work, we have studied the host galaxies of low-redshift quasars using our spectrophotometric decomposition technique.  We selected a sample of bright type 1 quasars with extended optical morphologies at $0.1<z<1$ from DESI. Leveraging the combination of high-quality HSC imaging and high-resolution DESI spectra, we applied our decomposition technique to 1083 quasars. The technique features joint modeling of the panchromatic spectral and photometric observations of the quasars. Mock tests demonstrate that our pipeline can break the degeneracy between AGN and host stellar components, and makes reliable estimation for their physical properties. 

Our results show that the above DESI quasar sample covers a large range of AGN luminosity, $M_{\mathrm{BH}}$, and host stellar mass $M_{\star}$. Absorption features were clearly observed in the decomposed host galaxy spectra. This quasar population exhibits low SFRs in their host galaxies and a high fraction (23\%) of post-starburst hosts. The majority of their host galaxies are quiescent, and some of them have old stellar populations. 
Furthermore, after incorporating systematic errors, the $M_{\mathrm{BH}} - M_{\star}$ scaling relation of our quasars is broadly consistent with the canonical local relations. We found no apparent correlation between $M_{\mathrm{BH}}$ and host velocity dispersion $\sigma_{\star}$ in our sample, which is primarily driven by the large measurement uncertainties at low spectral S/N.

Similar studies in the literature often focused on compact quasars and found that their host galaxies are usually star-forming galaxies. Our work is complementary to these previous studies and found that the host galaxies of extended quasars may have quite different properties, and possibly in different evolutionary stages. Future studies based on large-area space observations will provide more information on quasar host galaxies.

\section{Acknowledgements}

We acknowledge support from the National Key R\&D Program of China (2021YFA1600404) and the National Science Foundation of China (12225301). 
M.S. acknowledges support by the State Research Agency of the Spanish Ministry of Science and Innovation under the grants `Galaxy Evolution with Artificial Intelligence' (PGC2018-100852-A-I00) and `BASALT' (PID2021-126838NB-I00) and the Polish National Agency for Academic Exchange (Bekker grant BPN/BEK/2021/1/00298/DEC/1). This work was partially supported by the European Union's Horizon 2020 Research and Innovation program under the Maria Sklodowska-Curie grant agreement (No. 754510).
S.P. is supported by the international Gemini Observatory, a program of NSF NOIRLab, which is managed by the Association of Universities for Research in Astronomy (AURA) under a cooperative agreement with the U.S. National Science Foundation, on behalf of the Gemini partnership of Argentina, Brazil, Canada, Chile, the Republic of Korea, and the United States of America.

This material is based upon work supported by the U.S. Department of Energy (DOE), Office of Science, Office of High-Energy Physics, under Contract No. DE–AC02–05CH11231, and by the National Energy Research Scientific Computing Center, a DOE Office of Science User Facility under the same contract. Additional support for DESI was provided by the U.S. National Science Foundation (NSF), Division of Astronomical Sciences under Contract No. AST-0950945 to the NSF’s National Optical-Infrared Astronomy Research Laboratory; the Science and Technology Facilities Council of the United Kingdom; the Gordon and Betty Moore Foundation; the Heising-Simons Foundation; the French Alternative Energies and Atomic Energy Commission (CEA); the National Council of Humanities, Science and Technology of Mexico (CONAHCYT); the Ministry of Science, Innovation and Universities of Spain (MICIU/AEI/10.13039/501100011033), and by the DESI Member Institutions: \url{https://www.desi.lbl.gov/collaborating-institutions}. Any opinions, findings, and conclusions or recommendations expressed in this material are those of the author(s) and do not necessarily reflect the views of the U. S. National Science Foundation, the U. S. Department of Energy, or any of the listed funding agencies.
The authors are honored to be permitted to conduct scientific research on Iolkam Du’ag (Kitt Peak), a mountain with particular significance to the Tohono O’odham Nation.

\bibliography{references}
\bibliographystyle{aasjournal}

\appendix
\renewcommand{\thefigure}{A\arabic{figure}} 
\setcounter{figure}{0}

\section{Additional Imaging Decomposition Example}
\label{sec:appendix_A}
\myrev{In this appendix, we provide an additional example of our 2D imaging decomposition using \texttt{GalfitM}. This illustrates the separation of the central quasar point source from the extended host galaxy emission, supporting the use of these photometric priors in our spectral analysis.}

\begin{figure*}[htbp]
    \centering
    \includegraphics[width=0.8\linewidth]{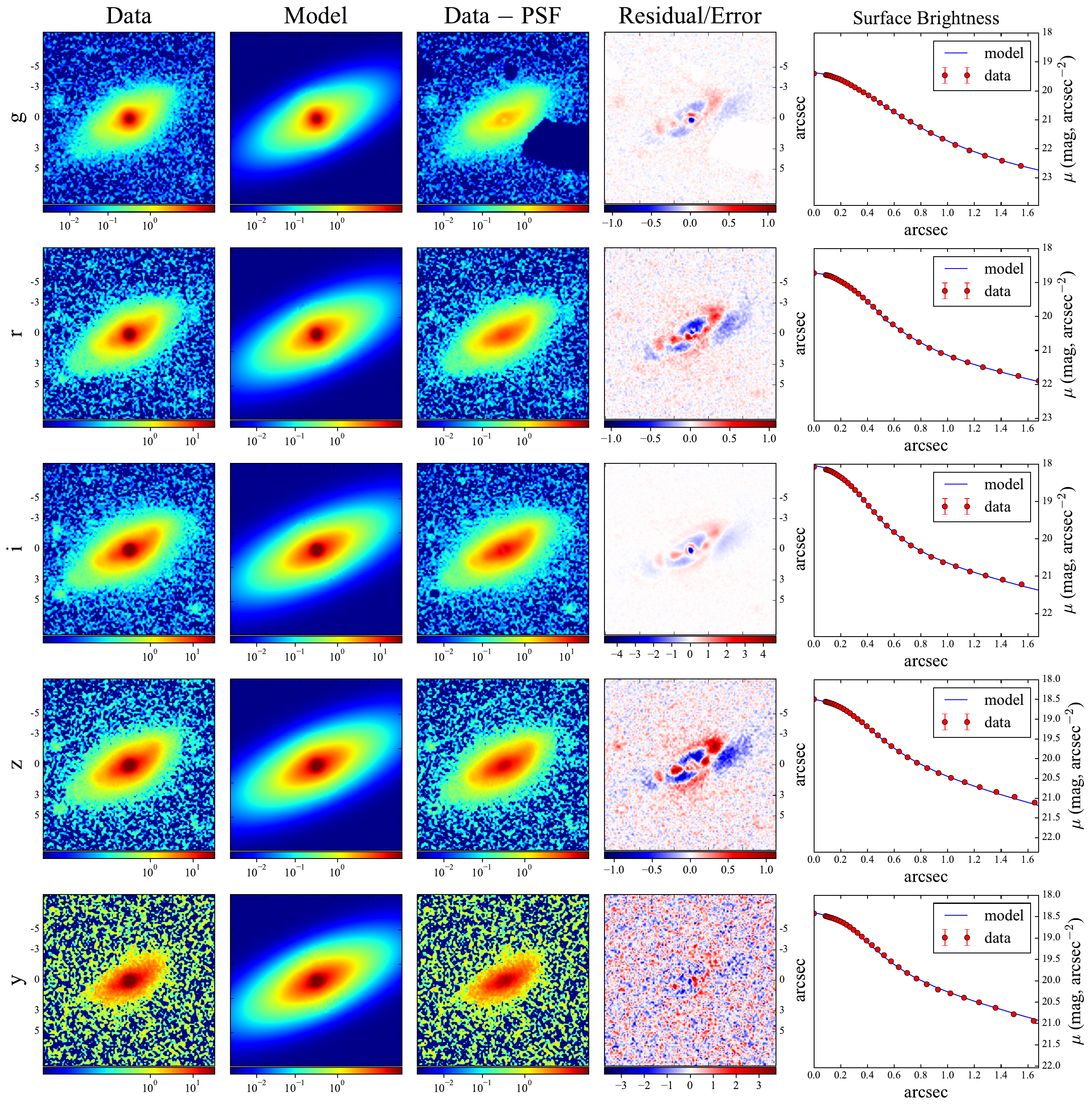}
    \caption{\myrev{\texttt{GalfitM} fitting results for a quasar at $z = 0.24$. Notations are the same as in Figure \ref{fig:galfitm_res1}.}}
    \label{fig:galfitm_res2}
\end{figure*}

\renewcommand{\thefigure}{B\arabic{figure}} 
\setcounter{figure}{0}

\section{Additional Mock Test Results}
\label{sec:appendix_B}
\myrev{In this appendix, we present supplementary statistical results from our mock tests to further validate the reliability of our spectrophotometric decomposition pipeline. Specifically, we display the recovery of the host galaxy median and mean flux levels as a function of the intrinsic input flux. This detailed comparison confirms that our decomposition approach successfully recovers the host galaxy properties without introducing severe systematic offsets, particularly for targets with non-negligible host contributions.}

\begin{figure}
    \centering
    \includegraphics[width=0.7\linewidth]{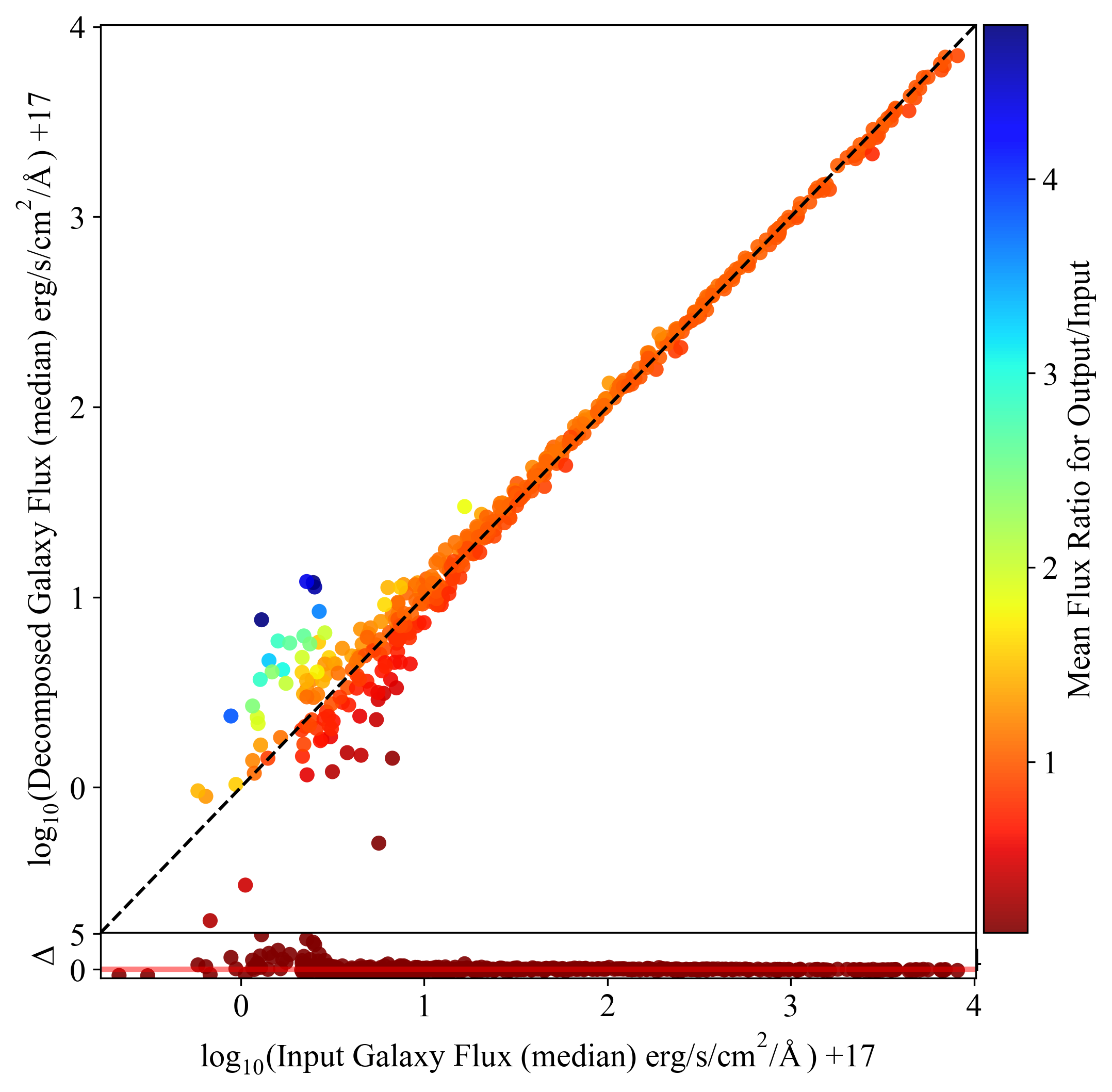}
    \caption{Recovered host galaxy flux levels for the second mock dataset as a function of the median flux of the input galaxy component, color-coded by the mean flux ratio between the decomposed galaxy spectra and the input ones. 
    There is a clear trend of our decomposed results converging to the one-to-one relation with their intrinsic values at higher host contribution. The attached lower panel shows the relative difference $\Delta = f_{\mathrm{out}}/f_{\mathrm{in}}-1$, with the reference red line indicating zero difference.}
    \label{fig:flux level recovery}
\end{figure}

\end{document}